\documentclass[aps,prb,reprint,nofootinbib,superscriptaddress]{revtex4-2}

\usepackage{amsmath,amssymb,amsthm,bm,physics,mathtools}
\usepackage{graphicx}
\usepackage{hyperref}

\begin{document}

\title{Dynamic Pseudogap Model}

\author{E.Z. Kuchinskii and M.V. Sadovskii}

\affiliation{Institute for Electrophysics, Russian Academy of Sciences,
Ural Branch,\\
Amundsen str. 106, Ekaterinburg 620016, Russia\\
E-mail: sadovski@iep.uran.ru}


\begin{abstract}

We formulate a theory of pseudogap formation generated by
dynamic finite nesting vector ${\bf Q}$ fluctuations with
characteristic oscillation frequency $\omega_0$, and damping $\gamma$.
Starting from a Hamiltonian describing electrons coupled to a classical
Gaussian random field, we introduce the double-series
representations of the single-particle Green's function within an
Abelian (commuting) approximation to the exact SU(2) time evolution
of the pseudogap problem.
The resulting propagator naturally acquires a generalized
Bogoliubov structure in which every stochastic scattering history is
characterized by an effective dynamic gap, leading to a coherent
superposition of dynamically broadened sidebands with complex Poisson
weights.

A central result of the theory is the emergence of a dynamically
generated decoherence scale $\Gamma_{\rm eff}$ governing the crossover
between two qualitatively different pseudogap regimes.
For $\omega_0>\Gamma_{\rm eff}$ the fluctuating field is resolved
coherently and the double-series representation provides a controlled
description of dynamic sideband formation.
Conversely, when $\Gamma_{\rm eff}\gtrsim\omega_0$, coherence is
progressively lost and the theory crosses over to the quasistatic
fluctuating-gap regime described by the exact continued-fraction
solution.
The coherent and quasistatic descriptions are therefore interpreted as
two complementary asymptotic limits of the same microscopic dynamic
pseudogap model.
The detailed results of numerical calculations for electron spectral density
and density of states supporting our approach are presented for different sets 
of model parameters confirming the general picture of this crossover.

\end{abstract}

\maketitle


\section{Introduction}

The properties of electrons in metals interacting with strong
short-range order fluctuations have attracted continuous interest for
several decades.
Examples include charge-density-wave (CDW), spin-density-wave (SDW) and excitonic
fluctuations, fluctuating Peierls order \cite{Gruner}, as well as the pseudogap
phenomena observed in underdoped cuprate superconductors
\cite{Timusk, ufn1, Tamm}.

A number of exactly (or nearly exactly) solvable models for one-dimensional
electrons interacting with pseudogap were developed based on the complete
diagrammatic series generated by the static Gaussian random field, which can
be summed exactly in the form of continued-fraction recursion relations.
These solutions constitute the basis of the fluctuating-gap model and
provide a remarkably accurate description of pseudogap formation in systems
with finite correlation length \cite{SadovskiiBook,Sad1,Sad2,Won1,Sad3,SadTim}.

These models were later generalized to two-dimensional case with special
emphasis on pseudogap physics in cuprates and other strongly correlated
systems \cite{PS1,PS2,SK98,PRB05,PRB07,ufn2,Chub_1,Chub_2}.

The static theory, however, neglects the intrinsic dynamics of the
fluctuating order parameter. Near structural or electronic phase transitions
the characteristic fluctuation frequency may become comparable to electronic
energy scales, making temporal fluctuations an essential part of the problem.
A systematic extension of the fluctuating-gap formalism to finite temporal
correlation time therefore represents an important open problem.

Recently we obtained \cite{KS2024} an exact solution for electrons
interacting with a spatially uniform ($Q=0$) Gaussian random field with
finite temporal correlations, generalizing the earlier dynamic reformulation
of exactly solvable Keldysh model \cite{SadovskiiBook} introduced by Kikoin
and Kiselev \cite{KK,EK}. Using generalized Ward identities we derived exact
recursion relations for the Green's function and demonstrated that the solution
admits an explicit double-series representation describing coherent dynamic
sidebands.

The present work addresses the considerably more complicated problem of
dynamic pseudogap fluctuations with a finite scattering vector ${\bf Q}$.
Unlike the $Q=0$ case, every scattering event transfers an electron
between the nested regions of the Fermi surface, producing an alternating
sequence of propagators with dispersions $\xi_{\bf k}$ and $\xi_{\bf k+Q}$,
giving rise to an effective SU(2) pseudospin structure absent in the uniform 
problem.

Our primary goal is to construct an analytical description of this dynamic
finite-$Q$ problem within the physically important regime where the
fluctuating order parameter may be regarded as a classical Gaussian random
field with finite temporal correlation time.
Employing an exact cumulant resummation of the commuting (Abelian) part of the
time-evolution operator, we derive explicit double-series Green's functions
for both commensurate and incommensurate fluctuations.

A central result of the present work is that every dynamic scattering
history may be associated with an effective complex gap parameter.
This naturally leads to a generalized Bogoliubov representation in which the
full Green's function is expressed as a coherent superposition of dynamically
broadened quasiparticle branches of Bogoliubov quasiparticles, producing
spectral densities with dynamically generated sidebands and densities of states
with dynamically induced modulation.

The coherent dynamic theory is shown to possess a well-defined domain of
applicability controlled by the competition between the characteristic
oscillation frequency of the fluctuating field and the dynamically
generated decoherence rate.
When coherent sidebands overlap strongly, the double-series expansion
gradually loses convergence and the calculated spectra evolve toward those
described by static continued fractions.
The static fluctuating-gap theory may be interpreted as the decohered infrared
limit of the more general dynamic theory developed here.

\section{Microscopic Model}

We consider electrons interacting with a fluctuating order parameter
characterized by a finite ordering wave vector ${\bf Q}$. Such a model describes
electronic scattering from short-range charge-density-wave (CDW),
spin-density-wave (SDW), excitonic, or Peierls fluctuations.
Throughout the present work the fluctuating field is assumed to be
classical, while the electronic subsystem is treated fully quantum
mechanically.

The electronic Hamiltonian is written as
\begin{equation}
H(t)=H_0+H_{\mathrm{int}}(t),
\label{FullHamiltonian}
\end{equation}
where
\begin{equation}
H_0=
\sum_{\mathbf{k}}
\xi_{\mathbf{k}}
c_{\mathbf{k}}^\dagger
c_{\mathbf{k}}
\label{BareHamiltonian}
\end{equation}
describes noninteracting electrons measured relative to the chemical
potential.

The interaction with the fluctuating order parameter is
\begin{equation}
H_{\mathrm{int}}(t)
=
\sum_{\mathbf{k}}
\left[
\Delta(t)
c_{\mathbf{k+Q}}^\dagger
c_{\mathbf{k}}
+
\Delta^*(t)
c_{\mathbf{k}}^\dagger
c_{\mathbf{k+Q}}
\right],
\label{InteractionHamiltonian}
\end{equation}
where $\Delta(t)$ denotes a spatially uniform but temporally fluctuating
complex random field.

Equation (\ref{InteractionHamiltonian}) describes repeated scattering
between two nested regions of the Fermi surface separated by the ordering
vector ${\bf Q}$. Every interaction event transfers an electron according to
$\mathbf{k}\longleftrightarrow
\mathbf{k+Q}$,
which constitutes the essential difference between the present problem and
the previously studied $Q=0$ dynamic model.

In most part of the paper we consider one-dimensional (Peierls-like CDW) system,
though generalization for two-dimensional case (scattering between hot-spots
on 2d Fermi surface) in principle is straightforward \cite{Tamm,PS1,PS2,SK98}).

An alternative schematic derivation of our approach based on Keldysh technique 
is presented in Appendix \ref{app:Keld}.

\subsection{Commensurate and Incommensurate Fluctuations}

The physical nature of the fluctuating field (order parameter) depends on the
symmetry of the underlying ordered state.

For commensurate ordering one may choose the order parameter to be real,
\begin{equation}
\Delta(\mathbf{r},t)
=
2\Delta(t)
\cos(\mathbf{Q}\cdot\mathbf{r}),
\label{CommensurateField}
\end{equation}
so that $\Delta(t)=\Delta^*(t)$.

For incommensurate order the order parameter is complex,
\begin{eqnarray}
\Delta(\mathbf{r},t)
=
\Delta(t)
e^{i\mathbf{Q}\cdot\mathbf{r}}
+
\Delta^*(t)
e^{-i\mathbf{Q}\cdot\mathbf{r}}=\nonumber\\
=2\Delta\cos(\mathbf{Q}\cdot\mathbf{r}+\phi)
\label{IncommensurateField}
\end{eqnarray}
where the complex amplitudes describe independent fluctuations of the two
components of the order parameter, $\Delta$ and $\phi$ are the modulus
and the phase of the order parameter, both random.

The distinction between these two cases becomes essential in the dynamic
problem because the individual scattering histories acquire different
effective gap parameters, leading to different structure of Green's functions.

More details on commensurate and incommensurate fluctuations are discussed
in Appendix \ref{app:Incomm_vs_Comm}.

\subsection{Gaussian Statistics}

The fluctuating field is assumed to obey Gaussian statistics with zero
average,
\begin{equation}
\langle\Delta(t)\rangle=0,
\end{equation}
and correlation function
\begin{equation}
\langle
\Delta(t)\Delta^*(0)
\rangle
=
\Delta^2
e^{-\gamma|t|}
\cos({\omega_0 t}).
\label{CorrelationFunction}
\end{equation}
Here
\begin{itemize}
\item $\Delta$ determines the characteristic fluctuation amplitude,
\item $\gamma^{-1}$ is the temporal correlation time,
\item $\omega_0$ denotes the characteristic oscillation frequency of the
dynamic order parameter.
\end{itemize}
Equation (\ref{CorrelationFunction}) represents the simplest stationary
random process supplemented by coherent oscillations.
The exponential form is particularly convenient because repeated time
integrations generated by the diagrammatic expansion may be evaluated
analytically.

For incommensurate order one additionally has
\begin{equation}
\langle
\Delta(t)\Delta(0)
\rangle
=
0,
\qquad
\langle
\Delta^*(t)\Delta^*(0)
\rangle
=
0,
\label{ComplexStatistics}
\end{equation}
whereas for commensurate fluctuations the field is real and only a single
independent correlator remains.

\subsection{Nambu Representation}

Throughout most of this paper we consider the vicinity of a nested part
(hot-spots) of the Fermi surface satisfying
$\xi_{\mathbf{k+Q}}\simeq-\xi_{\mathbf{k}}\equiv-\xi$.
so that every scattering event changes the sign of the electronic dispersion
$\xi\longrightarrow-\xi\longrightarrow\xi\longrightarrow\cdots$,
which will simplify the subsequent analysis. For incommensurate fluctuations
we neglect the states with ${\bf k}\pm 2{\bf Q}$ and higher momentum transfers
and use the two-patch linearized ($Q\approx 2p_F$) approximation.
Unlike the uniform ($Q=0$) \cite{KS2024} problem, successive scattering events
therefore alternate between two branches of the electronic spectrum.
This alternating structure is ultimately responsible for the generalized
Bogoliubov -- Gor'kov form of the dynamic Green's function derived below.

It is convenient to introduce the two-component Nambu spinor \cite{Nambu1960}
\begin{equation}
\hat\Psi_{\mathbf{k}}
=
\begin{pmatrix}
c_{\mathbf{k}}
\\
c_{\mathbf{k+Q}}
\end{pmatrix},
\label{Nam_spin}
\end{equation}
in terms of which Eq.~(\ref{FullHamiltonian}) assumes the compact form
\begin{equation}
H(t)
=
\sum_{\mathbf{k}}
\hat\Psi_{\mathbf{k}}^\dagger
\hat H(t)
\hat\Psi_{\mathbf{k}},
\label{Ham_Nam}
\end{equation}
with
\begin{equation}
\hat H(t)
=
\xi
\sigma_3
+
\Delta(t)\sigma_+
+
\Delta^*(t)\sigma_-,
\label{NambuHamiltonian}
\end{equation}
where
\begin{equation}
\sigma_\pm
=
\frac12
(\sigma_1\pm i\sigma_2)
\end{equation}
are the standard pseudospin raising and lowering operators.

For a commensurate order parameter Eq.~(\ref{NambuHamiltonian}) reduces to
\begin{equation}
\hat H(t)
=
\xi\sigma_3
+
\Delta(t)\sigma_1,
\end{equation}
whereas for incommensurate fluctuations both $\sigma_+$ and $\sigma_-$
components (fluctuations) remain independent.

The finite-$Q$ problem therefore possesses an intrinsic SU(2)
pseudospin structure originating entirely from the repeated alternation
between the nested electronic branches.
This noncommutative structure will be shown to generate the certain Magnus
corrections absent in $Q=0$ model.

\subsection{Exact Time Evolution and SU(2) Structure}

The retarded Green's function of a quantum particle interacting with dynamic
external field is defined by an equation:
\begin{equation}
\left(i\frac{\partial}{\partial t}-\hat H(t)\right)\hat G^R(t,t')=\delta(t-t')
\label{Gr_part}
\end{equation} 
Its solution is:
\begin{equation}
\hat G^R(t,t')=-i\theta(t-t')U(t,t')
\label{GRt}
\end{equation}
where time-evolution operator is given by:
\begin{equation}
U(t,t')=T\exp\left[-i\int_{t'}^{t}\hat H(\tau)d\tau\right]
\label{Utt}
\end{equation}
where $T$ denotes chronological ordering.
In case of the Gaussian random external field $\Delta(t)$ we introduce
\begin{equation}
\hat G^R(t)
=
-i\theta(t)\left<
T
\exp
\left[
-i
\int_0^t
dt'
\hat H(t')
\right]
\right>_{\Delta},
\label{ExactGreen}
\end{equation}
where we perform the averaging over this random field taking into account
Eq.~(\ref{CorrelationFunction}).

Equation (\ref{ExactGreen}) is formally exact.
The principal difficulty of the dynamic finite-$Q$ problem originates from
the noncommutativity of the Hamiltonian at different times,
\begin{equation}
[\hat H(t_1),\hat H(t_2)]
\neq
0,
\end{equation}
which prevents a direct evaluation of the time-ordered exponential.

Using Eq.~(\ref{NambuHamiltonian}) after straightforward algebra
one immediately obtains
\begin{align}
[\hat H(t_1),\hat H(t_2)]
&=
2\xi
\left[
\Delta(t_2)-\Delta(t_1)
\right]
\sigma_+
\nonumber\\
&\quad
-
2\xi
\left[
\Delta^*(t_2)-\Delta^*(t_1)
\right]
\sigma_-
\nonumber\\
&\quad
+
\left[
\Delta(t_1)\Delta^*(t_2)
-
\Delta(t_2)\Delta^*(t_1)
\right]
\sigma_3 .
\label{GeneralCommutator}
\end{align}
Equation (\ref{GeneralCommutator}) demonstrates that the noncommutativity
does not originate from quantum fluctuations of the order parameter.
The fluctuating field is purely classical.
Instead, it reflects the non-Abelian SU(2) pseudospin algebra associated
with repeated scattering between the two nested electronic branches.
The commutator vanishes identically only in special limiting cases.

The problem arising here is very similar to the well known difficulties in
construction diagrammatic perturbation theory for spin-dependent Hamiltonians
in the theory of magnetism. Noncommutativity of spin operators requires the
reformulation of the standard Wick theorem and leads to rather complicated
variants of diagram technique as proposed in Refs. \cite{VLP,IK,IS,BKY,BL}.
Below we follow rather simplified approach neglecting pseudospin structure of
the vertices in our theory.

\subsection{Commuting (Abelian) Limit}

The simplest case corresponds to an electron at the Fermi surface, $\xi=0$.
For commensurate (real order parameter $\Delta(t)$) fluctuations,
\begin{equation}
\Delta(t)=\Delta^*(t),
\end{equation}
the Hamiltonian becomes
\begin{equation}
\hat H(t)
=
\Delta(t)\sigma_1,
\end{equation}
so that
\begin{equation}
[\hat H(t_1),\hat H(t_2)]
=
0.
\end{equation}
Chronological ordering therefore becomes unnecessary,
\begin{equation}
T
\exp
\left[
-i
\int_0^t
dt'
\hat H(t')
\right]
=
\exp
\left[
-i
\int_0^t
dt'
\hat H(t')
\right].
\end{equation}
Consequently the dynamic problem reduces to the evaluation of an ordinary
Gaussian average and admits an exact cumulant resummation.
This observation forms the basis of the double-series representation
derived below.

For incommensurate fluctuations (complex order parameter $\Delta$) it is not so
in general as is clearly seen from (\ref{GeneralCommutator}) as  commutator
remains finite even at $\xi=0$.

Away from the Fermi surface in commensurate case
$\xi\neq0$, and for incommensurate case
in general, the Hamiltonians at different times no longer commute.

The evolution operator may formally be written in the Magnus form  \cite{Magnus}:
\begin{equation}
U(t)
=
\exp
\left[
\Omega_1
+
\Omega_2
+
\Omega_3
+\cdots
\right],
\label{MagnusExpansion}
\end{equation}
where
\begin{equation}
\Omega_1
=
-i
\int_0^t
dt'
\hat H(t'),
\end{equation}
while the leading correction is
\begin{equation}
\Omega_2
=
-\frac12
\int_0^t
dt_1
\int_0^{t_1}
dt_2
[\hat H(t_1),\hat H(t_2)].
\label{SecondMagnus}
\end{equation}
The higher Magnus terms describe successive SU(2) rotations of the
electronic pseudospin generated by the alternating sequence of
finite-$Q$ scattering events.

The finite-$Q$ problem therefore separates naturally into two
conceptually distinct ingredients.

The first is the Gaussian stochastic dynamics of the classical order
parameter, completely determined by the correlation function
(\ref{CorrelationFunction}).

The second is the noncommutative SU(2) evolution of the electronic
pseudospin induced by repeated transitions between the nested branches of
the Fermi surface. Only the latter is responsible for the Magnus corrections.

Throughout most of the present work we analyze the commuting sector.
Actually our numerical calculations show that for a wide and interesting region 
of the parameters of our model we can identify the parameter region in which 
the coherent double-series solution provides a reasonable description of dynamic 
pseudogap physics, while outside this region we can successfully apply another 
(continued fraction) approximation.

More details on physics of Magnus series effects can be found in
Appendix \ref{sec:SU2}.

\section{Exact Cumulant Solution in the Commuting Limit}
\label{sec:ExactCuml}

As discussed in the previous section, the full finite-$Q$ problem is
generally non-Abelian owing to the noncommutativity of the Hamiltonian at
different times.
However, in special case of electron on the Fermi surface for commensurate
fluctuations, the Hamiltonians commute and the time-ordered exponential in
Eq.~(\ref{ExactGreen}) simplifies considerably.

Actually the same approximation becomes exact also for the case of generalized
Keldysh model, corresponding to $Q=0$ \cite{KS2024}.

In general case we shall consider commuting approximation as a reasonable first
step also in general case and try to determine its region of applicability,
when the simple cumulant expansion provides accurate enough description of
our model.

\subsection{Time-Domain Green's Function}

Consider first the $Q=0$ problem with real $\Delta(t)$ when the retarded 
Green's function becomes
\begin{equation}
G^R(t)
=
-i
\theta(t)
e^{-i\xi t}
\left<
\exp
\left[
-i
\int_0^t
dt'
\Delta(t')
\right]
\right>.
\label{TimeGreen0}
\end{equation}
Since the fluctuating field obeys Gaussian statistics, all cumulants
higher than second order vanish identically.
The Gaussian average is therefore evaluated exactly by the standard
cumulant theorem \cite{Kubo_cum},
\begin{equation}
\left<
e^{A}
\right>
=
\exp
\left(
\frac12
\left<
A^2
\right>
\right),
\end{equation}
with
\begin{equation}
A
=
-i
\int_0^t
dt'
\Delta(t').
\end{equation}
Consequently
\begin{equation}
G^R(t)
=
-i
\theta(t)
e^{-i\xi t}
e^{-K(t)},
\label{TimeGreenCumulant}
\end{equation}
where the cumulant function is
\begin{equation}
K(t)
=
\frac12
\int_0^t
dt_1
\int_0^t
dt_2
\,
\langle
\Delta(t_1)
\Delta(t_2)
\rangle .
\label{GeneralCumulant}
\end{equation}
Equation (\ref{TimeGreenCumulant}) constitutes the Green's function
within the commuting approximation, which is actually exact in the $Q=0$ problem.

\subsection{Evaluation of the Cumulant and Emergent Dynamic Decoherence}
\label{SecGammaEff}

Using the exponential representation of our correlation function
\begin{equation}
\langle
\Delta(t)
\Delta(0)
\rangle
=
\frac{1}{2}\Delta^2
e^{-\gamma|t|}
[e^{-i\omega_0 t} + e^{i\omega_0 t}],
\label{expcorr}
\end{equation}
the double time integration in (\ref{GeneralCumulant}) may be performed
analytically, as shown below.

An important consequence of the cumulant representation is the natural
emergence of an effective dynamic decoherence rate governing the long-time
behavior of the electronic propagator.
Considering for shortness only the first term in (\ref{expcorr}) and introducing
$u=t_1-t_2$, one can write (\ref{GeneralCumulant}) as convolution
\begin{equation}
K(t)
=
\int_0^t
du
(t-u)
D(u).
\label{GammaK1}
\end{equation}
where
\begin{equation}
D(u)
=
\frac{1}{2}\Delta^2
e^{-\gamma u}
e^{-i\omega_0u},
\qquad
u>0,
\label{GammaK2}
\end{equation}
and the long-time limit may be analyzed by separating the cumulant into
time-independent and time-dependent parts.

Eq.~(\ref{GammaK1}) may be rewritten as
\begin{equation}
K(t)
=
t
\int_0^t
du\,D(u)
-
\int_0^t
du\,uD(u).
\label{GammaK3}
\end{equation}
For times much longer than correlation time,
$t\gg\gamma^{-1}$,
the upper integration limit may be extended to infinity with
exponentially small error,
\begin{equation}
K(t)
=
Bt
-
C
+
O(e^{-\gamma t}),
\label{GammaK4}
\end{equation}
where
\begin{equation}
B
=
\int_0^\infty
du\,
D(u),
\label{GammaB}
\end{equation}
and
\begin{equation}
C
=
\int_0^\infty
du\,
uD(u).
\end{equation}
The coefficient $B$ is readily evaluated,
\begin{equation}
B
=
\frac{1}{2}\Delta^2
\int_0^\infty
du\,
e^{-(\gamma+i\omega_0)u}
=
\frac{\Delta^2}
{2(\gamma+i\omega_0)}.
\label{GammaB2}
\end{equation}
Separating real and imaginary parts,
\begin{equation}
B
=
\frac{1}{2}\left[\frac{\gamma\Delta^2}
{\omega_0^2+\gamma^2}
-
i
\frac{\omega_0\Delta^2}
{\omega_0^2+\gamma^2}\right],
\label{GammaB3}
\end{equation}
one immediately identifies two physically distinct contributions.
The imaginary part produces a renormalization (shift) of the quasiparticle 
energy,
\begin{equation}
\delta\xi
=
-
\frac{1}{2}\frac{\omega_0\Delta^2}
{\omega_0^2+\gamma^2},
\label{Shift}
\end{equation}
whereas the real part generates irreversible decay of the propagator.
Accordingly we define the effective dynamic decoherence rate
\begin{equation}
\Gamma_{\mathrm{eff}}
=
\Re
\int_0^\infty
dt\,
\langle
\Delta(t)\Delta(0)
\rangle
=
\frac{\gamma\Delta^2}
{\omega_0^2+\gamma^2}.
\label{GammaEff}
\end{equation}
The contribution of the first term in (\ref{expcorr}) to the  Green's function
therefore assumes the asymptotic form
\begin{equation}
G(t)
\sim
e^{-i(\xi+\delta\xi)t}
e^{-\frac{1}{2}\Gamma_{\mathrm{eff}}t}
\label{GammaGF}
\end{equation}
Equation~(\ref{GammaEff}) demonstrates that
$\Gamma_{\mathrm{eff}}$ is not an additional phenomenological broadening
parameter but follows directly from the integrated temporal correlations.
It measures the irreversible loss of phase coherence caused by repeated
scattering from the dynamic pseudogap field.
The limiting behavior follows immediately from Eq.~(\ref{GammaEff}).

For slowly decaying fluctuations,
\begin{equation}
\gamma\ll\omega_0,
\end{equation}
one finds
\begin{equation}
\Gamma_{\mathrm{eff}}
\simeq
\frac{\gamma\Delta^2}{\omega_0^2},
\end{equation}
so that decoherence increases linearly with the inverse correlation time.

In the opposite motional narrowing limit,
\begin{equation}
\gamma\gg\omega_0
\end{equation}
the decoherence rate becomes
\begin{equation}
\Gamma_{\mathrm{eff}}
\simeq
\frac{\Delta^2}{\gamma},
\end{equation}
which decreases with increasing fluctuation rate.
Rapid fluctuations therefore average themselves out before electrons can
accumulate a significant scattering phase, restoring coherent
quasiparticle motion.

The quantity $\Gamma_{\mathrm{eff}}$ naturally determines the crossover
between coherent and incoherent pseudogap dynamics.
Indeed, the characteristic electronic coherence time is
\begin{equation}
\tau_\phi
=
\Gamma_{\mathrm{eff}}^{-1},
\end{equation}
whereas the characteristic oscillation period of the fluctuating field is
\begin{equation}
\tau_{\rm osc}
=
\omega_0^{-1}.
\end{equation}
Coherent sidebands remain well resolved provided
\begin{equation}
\tau_\phi
\gg
\tau_{\rm osc},
\end{equation}
or equivalently,
\begin{equation}
\omega_0
\gg
\Gamma_{\mathrm{eff}}.
\label{Criterion}
\end{equation}

When the opposite inequality holds,
\begin{equation}
\Gamma_{\mathrm{eff}}
\gg
\omega_0,
\end{equation}
electrons lose phase memory before resolving the intrinsic oscillations
of the fluctuating field.
The dynamics then becomes effectively quasistatic, explaining the
crossover toward the static fluctuating-gap description discussed in the
following sections.

Finally, it is worth emphasizing that the representation
(\ref{GammaEff}) is formally identical to the expressions governing
decoherence in Kubo stochastic line-shape theory and motional narrowing 
in magnetic resonance \cite{Kubo_res}.

Now, direct evaluation of the convolution integral with the account of both
signs of $\omega_0$ in (\ref{expcorr}) gives
\begin{align}
K(t)
&=
\frac{\Delta^2}
{2(\gamma+i\omega_0)^2}
\left[
(\gamma+i\omega_0)t
-
1
+
e^{-(\gamma+i\omega_0)t}
\right]
\nonumber\\
&
+
\frac{\Delta^2}
{2(\gamma-i\omega_0)^2}
\left[
(\gamma-i\omega_0)t
-
1
+
e^{-(\gamma-i\omega_0)t}
\right].
\label{EqCumExact}
\end{align}
The first terms inside the square brackets are linear in time
\begin{equation}
Bt
=
\left[\frac{1}{2}\Gamma_{\rm eff}\pm i\delta\xi\right]t,
\end{equation}
and therefore simply generate
\begin{equation}
e^{-K(t)}
=
e^{-\Gamma_{\rm eff}t}
e^{-K_{\rm ren}(t)},
\label{EqFactor}
\end{equation}
where 
\begin{equation}
K_{\rm ren}(t)
=
\lambda_1
\left[
1-e^{-\alpha_{+} t}
\right]
+
\lambda_2
\left[
1-e^{-\alpha_{-}t}
\right]
\label{K_red}
\end{equation}
where we have introduced
\begin{equation}
\alpha_{\pm}
=
\gamma
\pm
i\omega_0,
\end{equation}
and
\begin{equation}
\lambda_1
=
\frac{\Delta^2}
{2(\omega_0-i\gamma)^2},
\qquad
\lambda_2
=
\frac{\Delta^2}
{2(-\omega_0-i\gamma)^2}.
\label{lambi}
\end{equation}
Eq. (\ref{K_red}) contains only the bounded oscillatory part of the cumulant.
This procedure is entirely analogous to the standard decomposition of a
self-energy into an energy shift and damping rate.

\subsection{Double-Series Representation of the Dynamic Green's Function}

As shown below, the exponential structure of the cumulant naturally
generates a double Poisson series, in which every term corresponds to a
particular sequence of coherent dynamic scattering events.
Substituting (\ref{EqFactor}) and (\ref{K_red}) into
Eq.~(\ref{TimeGreenCumulant}) we get:
\begin{align}
G^R(t)
&=
-i\theta(t)
e^{-i\xi t}e^{-\Gamma_{\mathrm{eff}}t}
\nonumber\\
&\times
\exp(-\lambda_1-\lambda_2)
\exp
\left[
\lambda_1e^{-\alpha_+t}
+
\lambda_2e^{-\alpha_-t}
\right].
\label{GreenBeforeExpansion}
\end{align}
Expanding the second exponential here we immediately obtain
\begin{eqnarray}
G^R(t)
=
-i\theta(t)
e^{-i\xi t}e^{-\Gamma_{\mathrm{eff}}t}
e^{-\lambda_1-\lambda_2}
\times\nonumber\\
\times
\sum_{n,m=0}^{\infty}
W_{nm}
e^{-(n+m)\gamma t}
e^{-i(n-m)\omega_0t},
\label{FinalTimeSeries}
\end{eqnarray}
where
\begin{equation}
W_{nm}
=
\frac{\lambda_1^{\,n}\lambda_2^{\,m}}
{n!m!}.
\label{HistoryWeights}
\end{equation}
The Fourier transform of Eq.~(\ref{FinalTimeSeries}) is elementary:
\begin{equation}
G^R(\omega)
=
\int_0^\infty
dt\,
e^{i\omega t}
G^R(t),
\end{equation}
giving
\begin{widetext}
\begin{equation}
G^R(\omega,\xi)
=e^{-\lambda_1-\lambda_2}
\sum_{n,m=0}^{\infty}
W_{nm}
\frac{1}
{\omega-\xi
-(n-m)\omega_0
+i(n+m)\gamma + i\Gamma_{\mathrm{eff}}}.
\label{FrequencyDoubleSeries}
\end{equation}
\end{widetext}
where we have explicitly separated an additional decoherence
scattering rate $\Gamma_{\mathrm{eff}}$ which physical origin was discussed
above in \ref{SecGammaEff}.

Equation (\ref{FrequencyDoubleSeries}) is the exact coherent
double-series representation for $Q=0$ problem (when the commuting approximation
is exact). It can be written in convenient equivalent form as:
\begin{widetext}
\begin{equation}
G^R(\omega,\xi)
=\sum_{n,m=0}^{\infty}
P_{\lambda_1}(n)P_{\lambda_2}(m)
\frac{1}
{\omega-\xi
-(\omega_0-i\gamma)(n-\lambda_1)-(-\omega_0-i\gamma)(m-\lambda_2)},
\label{FreqDoubleSer}
\end{equation}
\end{widetext}
where we have introduced complex Poisson distributions:
\begin{equation}
P_{\lambda_i}(k)=\frac{\lambda_i^k}{k!}\exp(-\lambda_i),\, i=1,2
\label{Poiss}
\end{equation}
Actually, Eqs. (\ref{FrequencyDoubleSeries})-(\ref{FreqDoubleSer}) represent
an exact solution of Keldysh $Q=0$ model and the procedure described above
gives an alternative way to derive this result as compared with diagrammatic
approach of Ref. \cite{KS2024}.

\subsection{Dynamic Scattering Histories and Sidebands}

Each pair of integers $(n,m)$ labels one elementary dynamic history.
The integer $n$ counts the number of scattering events associated with the 
positive frequency component $e^{-i\omega_0t}$, while $m$
counts the corresponding number for the negative-frequency component
$e^{+i\omega_0t}$.
Consequently $n-m$ determines the net dynamical energy shift,
whereas $n+m$ governs the accumulated decoherence produced by the finite 
temporal correlation time.
The Green's function therefore appears as a coherent superposition of
infinitely many dynamically broadened sidebands.

Unlike ordinary Poisson probabilities, our distributions $P_{\lambda_i}(k)$ 
are in general complex 
containing both damping and oscillatory contributions.
Consequently, the individual histories cannot be interpreted as classical
probabilities.
Instead, they represent coherent amplitudes that interfere after summation
over all dynamic histories.
Only the complete Green's function possesses direct physical meaning.

From Eq. (\ref{FrequencyDoubleSeries}) it is obvious that
dynamic spectrum consists of an infinite family of
equidistant sidebands separated by the characteristic fluctuation
frequency.
The intensity of each sideband is determined by the corresponding complex
Poisson amplitude, while its width increases linearly with the total number
of scattering events.
The resulting structure resembles a Floquet spectrum.
However, unlike conventional Floquet systems generated by periodic driving,
the present sidebands arise entirely from stochastic Gaussian dynamics and
remain intrinsically broadened by the finite correlation time of the
fluctuating order parameter.

\section{Generalized Bogoliubov Representation of Dynamic Histories}

\subsection{History-dependent Nambu gaps}

For finite $Q$ problem each scattering event transfers an electron 
from one side of the Fermi surface to the other, or from one Nambu
branch to an another one. To generalize our $Q=0$ treatment to this
case we have to define the set of random matrix Poissonian fields which are 
non diagonal in Nambu representation. We must stress that our derivation is 
based on rather heuristic physical assumptions and is justified mostly by the 
results of our detailed numerical calculations presented below, which 
demonstrate the full agreement with the known exactly solvable limits of the
model under discussion. The rigorous derivation requires further studies, 
which we leave for the future. 

For commensurate fluctuations, starting from the obvious analogy with
exactly solvable $Q=0$ model,  we introduce two  matrix dynamic 
scattering fields as: 
\begin{eqnarray}
\hat\Delta_{1n}=\sigma_1\Delta_{1n}=\sigma_1(\omega_0-i\gamma)(n-\lambda_1),
\nonumber\\
\hat\Delta_{2m}=\sigma_1\Delta_{2m}=\sigma_1(-\omega_0-i\gamma)(m-\lambda_2),
\label{eq:delta12DS}
\end{eqnarray}
distributed according to complex Poisson distributions defined in (\ref{Poiss}).
The corresponding complex pseudogap scale is
\begin{equation}
\Delta_{nm}
=
\Delta_{1n}+\Delta_{2m}.
\label{eq:LambdaCommDS}
\end{equation}
so that the acting Poissonian field is
\begin{equation}
\hat\Delta_{nm}=\sigma_1\Delta_{nm}
=
\sigma_1(\omega_0-i\gamma)(n-\lambda_1)
+
\sigma_1(-\omega_0-i\gamma)(m-\lambda_2).
\label{DynamicGap}
\end{equation}
Note that 
\begin{equation}
\Delta_{nm}= 
(n-m)\omega_0
-i(n+m)\gamma - i\Gamma_{\mathrm{eff}}.
\label{Delta00}
\end{equation}
where we explicitly separated $i\Gamma_{\rm eff}$ contribution from
(\ref{DynamicGap}) where it is somehow hidden.
The sign of this imaginary quantity here is convention-dependent because 
it depends on the Fourier and retarded Green's function conventions. 

Now for a given history $(n,m)$ the effective Nambu matrix Hamiltonian
is assumed to take the following form:
\begin{equation}
\hat H_{nm}
=
\xi\sigma_3+\Delta_{nm}\sigma_1.
\label{HistoryHamiltonian}
\end{equation}
so that $\Delta_{nm}$ generates non-diagonal scattering in Nambu space producing
gaps. 

The corresponding $(n,m)$ history Green's function for finite $\xi$ is written 
as:
\begin{equation}
\hat G_{nm}(\omega,\xi)=[(\omega+i\Gamma_{\rm eff})\sigma_0
-\xi\sigma_3-\Delta_{nm}\sigma_1]^{-1}
\label{Gr_Numb}
\end{equation}
where we have explicitly added the effective damping introducing the
substitution $\omega\to\omega+i\Gamma_{\rm eff}$ in the diagonal part of the 
history Green's function. 
This is necessary to suppress unphysical contributions from $n=m$ terms,
reflected in the appearance of pathological discrete poles with both 
positive and negative weights contributions to spectral density. 
The precise value of this damping equal to $\Gamma_{\rm eff}$ is actually the 
minimal one sufficient to suppress these unphysical singularities.
This was explicitly checked by our numerical calculations.

Our numerical calculations also show that this substitution also ensures the 
full agreement with the results obtained for $\omega_0=0$ case from the exact
continued fraction representation to be discussed below in the limit of
large $\gamma$.
However, this substitution leads to inevitable double-counting of effective 
damping $\Gamma_{\rm eff}$ in case of $\xi=0$, where we actually have an exact 
solution, the same as in Keldysh model \cite{KS2024}.
To avoid this double counting in numerical calculations we have to redefine 
$\Delta_{nm}$ in commensurate case by subtracting $\Gamma_{\rm eff}$:  
$\Delta_{nm}\to (n-m)\omega_0-i(n+m)\gamma$.
We have explicitly checked numerically that such subtraction guarantees the
reproducibility of an exact solution of our model at $\xi=0$. We must stress
that this is to be done only in the diagonal part of $\hat G_{nm}$
to avoid double counting problem and restore the exact solution for
$\xi=0$. The physical definition of the dynamic gap $\Delta_{nm}$ remains that 
of Eq. (\ref{DynamicGap}), which in particular guarantees the absence of
average anomalous Green's function in our problem 
(see below Eq. (\ref{zero_gap})).

Finally we obtain the Nambu matrix representation of the single
$(n,m)$ history Green's function as:
\begin{equation}
\hat G_{nm}(\omega,\xi)
=
\frac{
(\omega+i\Gamma_{\rm eff})\sigma_0+\xi\sigma_3+\Delta_{nm}\sigma_1}
{(\omega+i\Gamma_{\rm eff})^2-\xi^2-\Delta_{nm}^2
}.
\label{eq:NambuResolventDS}
\end{equation}
The complex quasiparticle energy is defined as
\begin{equation}
E_{nm}
=
\sqrt{\xi^2+\Delta_{nm}^2},
\label{eq:EnmDS}
\end{equation}
The square root here must be assigned the retarded branch.
In the numerical implementation, this is implemented by first evaluating the
principal complex square root and then imposing
\begin{equation}
\operatorname{Im}E_{nm}\leq0,
\end{equation}
and
\begin{equation}
E_{nm}\rightarrow -E_{nm}
\qquad
\text{if}\qquad
\operatorname{Im}E_{nm}>0.
\label{eq:RetardedBranchDS}
\end{equation}
This branch prescription is essential for maintaining the retarded analytic
structure of each history.

\subsection{Bogoliubov decomposition}

The Nambu expression for the Green's function can alternatively be decomposed 
into Bogoliubov form. We define Bogoliubov coefficients as:
\begin{equation}
u_{nm}^2
=
\frac12
\left(
1+\frac{\xi}{E_{nm}}
\right),
\end{equation}
and
\begin{equation}
v_{nm}^2
=
\frac12
\left(
1-\frac{\xi}{E_{nm}}
\right).
\end{equation}
which satisfy
\begin{equation}
u_{nm}^2+v_{nm}^2=1.
\end{equation}
The component of the retarded propagator associated with one history 
is then represented as
\begin{equation}
G_{nm}^R(\omega,\xi)
=
\frac{u_{nm}^2}
{\omega+i\Gamma_{\rm eff}-E_{nm}}
+
\frac{(v_{nm}^2)^*}
{\omega+i\Gamma_{\rm eff}+E_{nm}^*}.
\label{eq:GnmPhysicalDS}
\end{equation}
Combining the Poisson histories with the Nambu matrix  gives the history 
averaged Green's function:
\begin{widetext}
\begin{equation}
G^R(\omega,\xi)
=
e^{-\lambda_1-\lambda_2}
\sum_{n,m}
\left[
W_{nm}
\frac{u_{nm}^2}
{\omega+i\Gamma_{\rm eff}-E_{nm}}
+W_{nm}^*
\frac{(v_{nm}^2)^*}
{\omega+i\Gamma_{\rm eff}+E_{nm}^*}
\right].
\label{eq:FullDoubleSeriesDS2}
\end{equation}
\end{widetext}
The first term here represents propagation through the positive-energy Nambu branch, 
whereas the second is the conjugated negative-energy branch. The conjugation of 
both the Poisson weight and the coherence factor is required to reproduce 
the retarded Green's function structure.

In the following for shortness  we use the following notation:
\begin{equation}
\tilde\omega=\omega\pm i\Gamma_{\rm eff},
\label{omega_xi}
\end{equation}
to fix the retarded or advanced behavior of the Green's function.

\subsection{Central result for the Green's function}

Thus the  double-series representation 
admits a clear formulation for our finite $Q$ problem.
Instead of interpreting every term as an independent dynamic sideband,
each scattering history may be regarded as an effective Bogoliubov
quasiparticle characterized by a complex dynamic gap parameter.
Near a nested portion of the Fermi surface the electronic states
$\mathbf{k}$ and $\mathbf{k\pm Q}$ become strongly mixed by the fluctuating
density-wave order parameter. We use Nambu notations (\ref{Nam_spin}),
(\ref{Ham_Nam}) and (\ref{NambuHamiltonian}) and work with corresponding
matrix Green's functions.

The essential complication of the dynamic finite-$Q$ problem is not the
Gaussian averaging itself but the noncommutativity of the SU(2) matrices
appearing at different scattering vertices.
The central approximation adopted in the present work consists in
projecting this ordered SU(2) algebra onto its commuting (Abelian)
sector. 

We associate with every stochastic history $(n,m)$ an effective 
Bogoliubov Hamiltonian (\ref{HistoryHamiltonian}).
The corresponding history Green's function is defined by $(1,1)$ matrix element
of the Nambu matrix (\ref{eq:NambuResolventDS}):
\begin{equation}
G_{nm}(\omega,\xi)
=
\frac{\tilde\omega+\xi}
{\tilde\omega^2-\xi^2-\Delta_{nm}^2},
\label{HistoryGreenFunction}
\end{equation}
which has exactly the Gor'kov--Bogoliubov form \cite{SadovskiiBook,Gorkov1958}.
It can be easily seen that for $\xi=0$ Eq. (\ref{HistoryGreenFunction}) 
reduces to the half-sum of two contributions, similar to those entering the 
$(n,m)$ - sum in Eq. (\ref{FreqDoubleSer}) corresponding to $Q=0$ problem and 
differing only by the signs of $\Delta_{nm}$.

Equation (\ref{HistoryGreenFunction}) constitutes the main working
approximation of the present theory. 
It is exact in the commuting limit ($\xi=0$ for commensurate fluctuations), 
reproduces the exact static Bogoliubov propagator when 
$\omega_0\rightarrow 0$ and  $\gamma\rightarrow 0$, and retains the correct 
free-electron limit $|\xi|\rightarrow\infty$. 
Away from the commuting limit, deviations originate 
entirely from the non-Abelian SU(2) vertex algebra neglected in the Abelian 
projection and will be  analyzed later using the Magnus expansion.
Actually this is the main approximation of the present work.

The excitation spectrum corresponding to one history is determined by
$E_{nm}=\sqrt{\xi^2+\Delta_{nm}^2}$.
Because $\Delta_{nm}$ is generally complex,
the quasiparticle energy also becomes complex,
with the imaginary part describing the finite lifetime generated by the
dynamic fluctuations.

Thus every history represents a dynamically broadened Bogoliubov
quasiparticle rather than a stationary excitation and the full Green's 
function is written as history averaged Gor'kov's function:
\begin{equation}
G(\omega,\xi)
=\sum_{n,m=0}^{\infty}
P_{\lambda_1}(n)P_{\lambda_2}(m)
\frac{\tilde\omega+\xi}
{\tilde\omega^2-\xi^2-\Delta_{nm}^2}
\label{GorFreqDoubleSer}
\end{equation}
However, we have to be cautious here to guarantee correct analytical
properties.
For accurate representation of retarded Green's function we have to use
Bogoliubov -- Gor'kov form:
\begin{widetext}
\begin{equation}
G^R(\omega,\xi)
=\sum_{n,m=0}^{\infty}
P_{\lambda_1}(n)P_{\lambda_2}(m)
\frac{u_{nm}^2}
{\omega+i\Gamma_{\rm eff}-\sqrt{\xi^2+\Delta_{nm}^2}}
+\sum_{n,m=0}^{\infty}
P_{\lambda_1}^*(n)P_{\lambda_2}^*(m)
\frac{(v^{2}_{nm})^*}
{\omega+i\Gamma_{\rm eff}+\sqrt{\xi^2+(\Delta^{2}_{nm})^*}}
\label{BogolFreqDoubleSer}
\end{equation}
\end{widetext}
Since $E_{nm}$ is complex, the coherence factors (Bogoliubov $u-v$ coefficients) 
themselves become complex amplitudes.
They therefore cannot be interpreted as occupation probabilities.
Instead they determine the relative weights of electron-like and hole-like
components within each coherent scattering history.
The complete Green's function is obtained by coherent summation over all
histories.
Equations (\ref{GorFreqDoubleSer}) and (\ref{BogolFreqDoubleSer}) constitute 
the central analytical result of the present work.

Unlike the conventional Bogoliubov theory, which contains only a single
gap parameter, the present dynamic problem generates an infinite family of
history-dependent complex gaps.
The physical Green's function emerges only after coherent summation over
all such histories.
The representation (\ref{BogolFreqDoubleSer}) admits a simple physical
interpretation.
The fluctuating order parameter generates an ensemble of dynamically
evolving pseudogaps rather than a single static gap.
Every coherent history experiences its own effective complex gap
(\ref{DynamicGap}), determined by the accumulated oscillatory phase and
decoherence acquired during the sequence of scattering events.
The observed quasiparticle spectrum therefore corresponds not to one
Bogoliubov branch but to the coherent superposition of infinitely many
dynamic Bogoliubov quasiparticles.
This viewpoint provides a natural bridge between the static
fluctuating-gap model and the coherent double-series representation
developed in the present work. Several important observations follow 
immediately.

First, the generalized Bogoliubov representation is exact within the
commuting approximation and introduces no additional assumptions beyond the
double-series expansion itself.

Second, the effective gap (\ref{DynamicGap}) is not an externally imposed
order parameter but an emergent quantity associated with a particular
dynamic scattering history.

The generalized Bogoliubov representation developed up to now
naturally describes commensurate pseudogap fluctuations, for which
the order parameter is real.
For incommensurate fluctuations the situation is qualitatively different.
In this case the fluctuating order parameter possesses two
independent fluctuating components and the corresponding history Green's
functions acquire a different structure.

\subsection{Incommensurate fluctuations}

For incommensurate fluctuations the order parameter is intrinsically
complex,
\begin{equation}
\Delta(t)
=
\Delta_1(t)
+
i\Delta_2(t),
\end{equation}
where the two real components fluctuate independently.
The corresponding Nambu Hamiltonian is
\begin{equation}
\hat H(t)
=
\xi\sigma_3
+
\Delta_1(t)\sigma_1
+
\Delta_2(t)\sigma_2.
\label{IncommHamiltonian}
\end{equation}
The square of this Hamiltonian is given by:     
\begin{equation}
H^2
=
\xi^2
+
\Delta_1^2
+
\Delta_2^2,
\end{equation}
so that the spectrum depends only upon the rotationally invariant
combination
\begin{equation}
\Delta^2
=
\Delta_1^2+\Delta_2^2.
\label{GapInvariant}
\end{equation}
Thus our fluctuating-gap model possesses an internal
$O(2)$ symmetry corresponding to rotations in the
$(\Delta_1,\Delta_2)$ plane in Nambu space.

Within the coherent cumulant expansion the Gaussian averages are again
performed exactly. Since
\begin{equation}
\langle
\Delta(t)\Delta(0)
\rangle
=
0,
\end{equation}
while
\begin{equation}
\langle
\Delta(t)\Delta^*(0)
\rangle
\neq0,
\end{equation}
the two Cartesian components in Nambu space generate two 
statistically independent stochastic histories.
Accordingly, the commensurate history scalar naturally generalizes to the
history vector in two-dimensional pseudospin (Nambu) space
\begin{equation}
\mathbf{\Delta}_{nm}
=
(\Delta_{1n},\Delta_{2m}),
\end{equation}
whose components are introduced as
\begin{equation}
\Delta_{1n}
=
(\omega_0-i\gamma)
(n-\lambda_1),
\label{Gap1}
\end{equation}
and
\begin{equation}
\Delta_{2m}
=
(-\omega_0-i\gamma)
(m-\lambda_2).
\label{Gap2}
\end{equation}
Now  the effective Nambu matrix Hamiltonian
for a given history $(n,m)$ is written as:
\begin{equation}
\hat H_{nm}
=
\xi\sigma_3+\Delta_{1n}\sigma_1+\Delta_{2m}\sigma_2.
\label{HistHamiltonian}
\end{equation}
and corresponding $(n,m)$ history Green's function is given by:
\begin{equation}
\hat G_{nm}(\omega,\xi)=[(\omega+i\Gamma_{\rm eff})\sigma_0
-\xi\sigma_3-\Delta_{1n}\sigma_1-\Delta_{2m}\sigma_2]^{-1}
\label{Gr_Nambu}
\end{equation}
The explicit introduction of effective damping in the denominator here again
ensures the numerical agreement with the results of an exact continued
fraction solution for the case of $\omega_0=0$ and in the limit of large
$\gamma$. There is no double counting here as $\Gamma_{\rm eff}$ actually does
not enter explicitly  our $\Delta_{1n}\sigma_1-\Delta_{2m}\sigma_2$
combination for incommensurate case.

The effective gap entering the history Green's function must preserve the
same $O(2)$ invariant in pseudospin space as in the static theory.
We therefore define
\begin{equation}
\mathbf{\Delta}_{nm}^2
=
\Delta_{1n}^2
+
\Delta_{2m}^2,
\label{DynamInvariant}
\end{equation}
which represents the natural dynamic generalization of the static gap
invariant (\ref{GapInvariant}).

Then the (1,1) matrix element of Nambu matrix for corresponding history 
Green's function is written as
\begin{equation}
G_{nm}(\omega,\xi)
=
\frac{\tilde\omega+\xi}
{\tilde\omega^2-\xi^2-
\mathbf{\Delta}_{nm}^2}.
\label{IncomHist}
\end{equation}
Equation (\ref{IncomHist}) constitutes the minimal
$O(2)$-symmetric extension of the commensurate coherent theory.
Rather than introducing a single effective complex history gap,
each stochastic history is characterized by a two-component history
vector whose invariant length determines the quasiparticle spectrum,
exactly as in the static fluctuating-gap model.

As in the commensurate theory, the resulting Green's function should be
viewed as the central working approximation of the coherent dynamic
theory. Its validity is supported by the exact static limit, the correct
free-electron limit, and the symmetry of the underlying Nambu
Hamiltonian, while corrections originating from the non-Abelian
structure of the exact time evolution are expected to appear through the
Magnus terms discussed below.

\subsection{Unified Physical Picture}

The proposed history Green's function possesses the correct static limit
for
\begin{equation}
\omega_0\rightarrow0,
\qquad
\gamma\rightarrow0,
\end{equation}
Explicit calculations for this limit are presented in Appendix \ref{app:Static}.
For incommensurate cases we reproduce the familiar result of
Refs. \cite{Sad1,Sad2} with Gor'kov's function averaged over Rayleigh
distribution of static gaps, while for commensurate case the averaging goes over
the usual Gaussian distribution \cite{Won1},

For finite values of $\omega_0$ the situation changes qualitatively.
Every scattering history acquires the energy shift
\begin{equation}
(n-m)\omega_0,
\end{equation}
together with the damping
\begin{equation}
(n+m)\gamma.
\end{equation}
The spectrum therefore consists of an infinite family of coherent
sidebands, whose amplitudes are determined by the complex Poisson weights.

Average gap is easily shown to be zero
\begin{eqnarray}
\left<{\Delta_{nm}}\right>=\sum_{n,m=0}^\infty P_{\lambda_1}(n)P_{\lambda_2}(m)
\times\nonumber\\
\times[(\omega_0-i\gamma)(n-\lambda_1)+(-\omega_0-i\gamma)(m-\lambda_2)]=0
\nonumber\\
\label{zero_gap}
\end{eqnarray}
corresponding to the absence of long-range order and vanishing anomalous
Green's functions.

The present analysis suggests that the dynamic pseudogap problem possesses
three physically distinct regimes.
\begin{enumerate}
\item
For
$\omega_0\gg\Gamma_{\rm eff}$,
the spectrum is governed by coherent dynamic sidebands described by the
double-series representation.
\item
For
$\omega_0\sim\Gamma_{\rm eff}$,
the sidebands overlap and coherence is progressively lost.
The calculated spectra evolve continuously toward the static ($\omega_0=0$)
continued-fraction solution.
\item
For
$\gamma\gg\omega_0,\Delta$
motional narrowing suppresses pseudogap formation altogether and restores
conventional free quasiparticle behavior.
\end{enumerate}
The static $\omega_0=0$ theory thus appears naturally as quasistatic limit of 
the more general dynamic theory.

\begin{figure*}
\includegraphics[clip=true,width=0.45\textwidth]{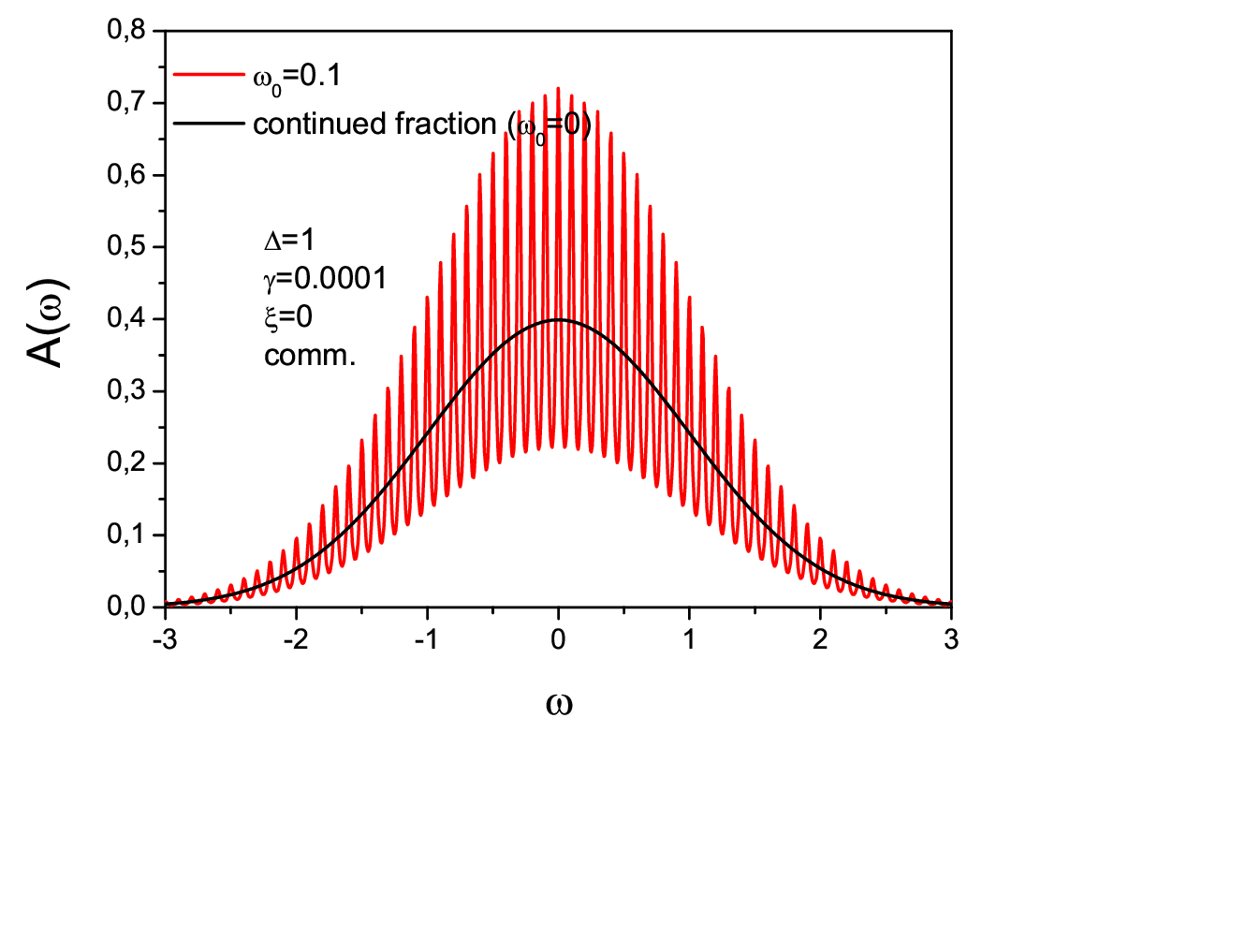}
\includegraphics[clip=true,width=0.45\textwidth]{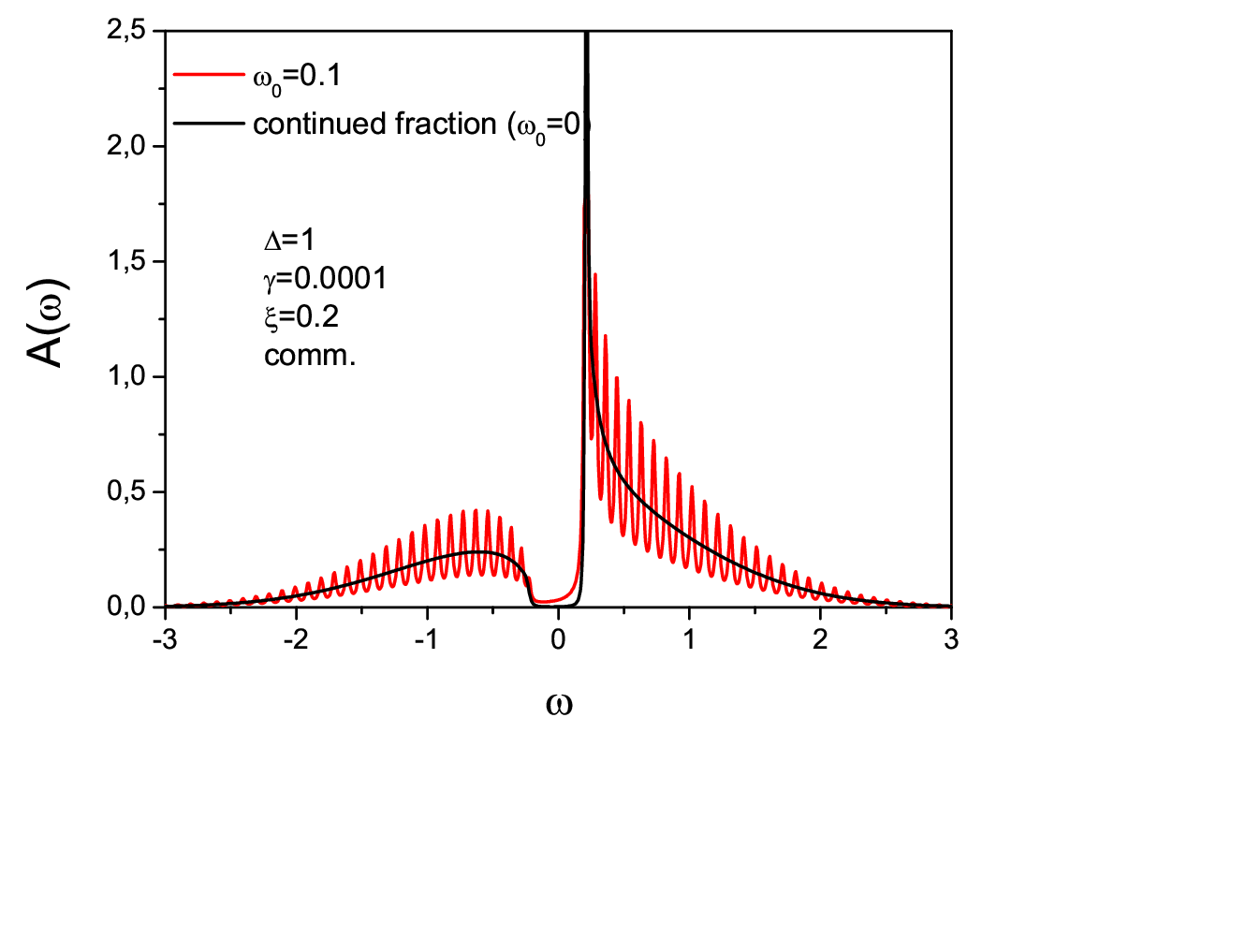}
\includegraphics[clip=true,width=0.45\textwidth]{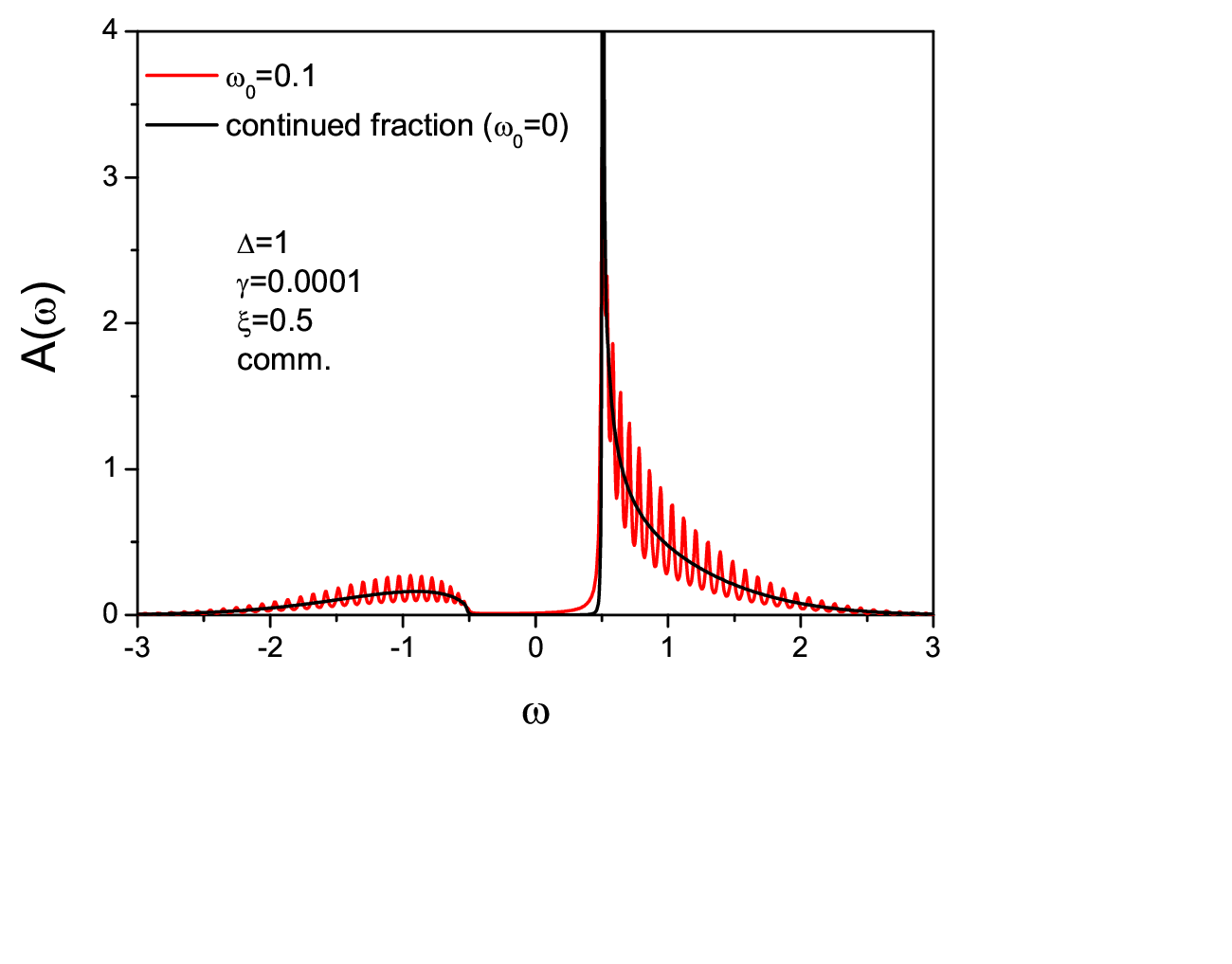}
\includegraphics[clip=true,width=0.45\textwidth]{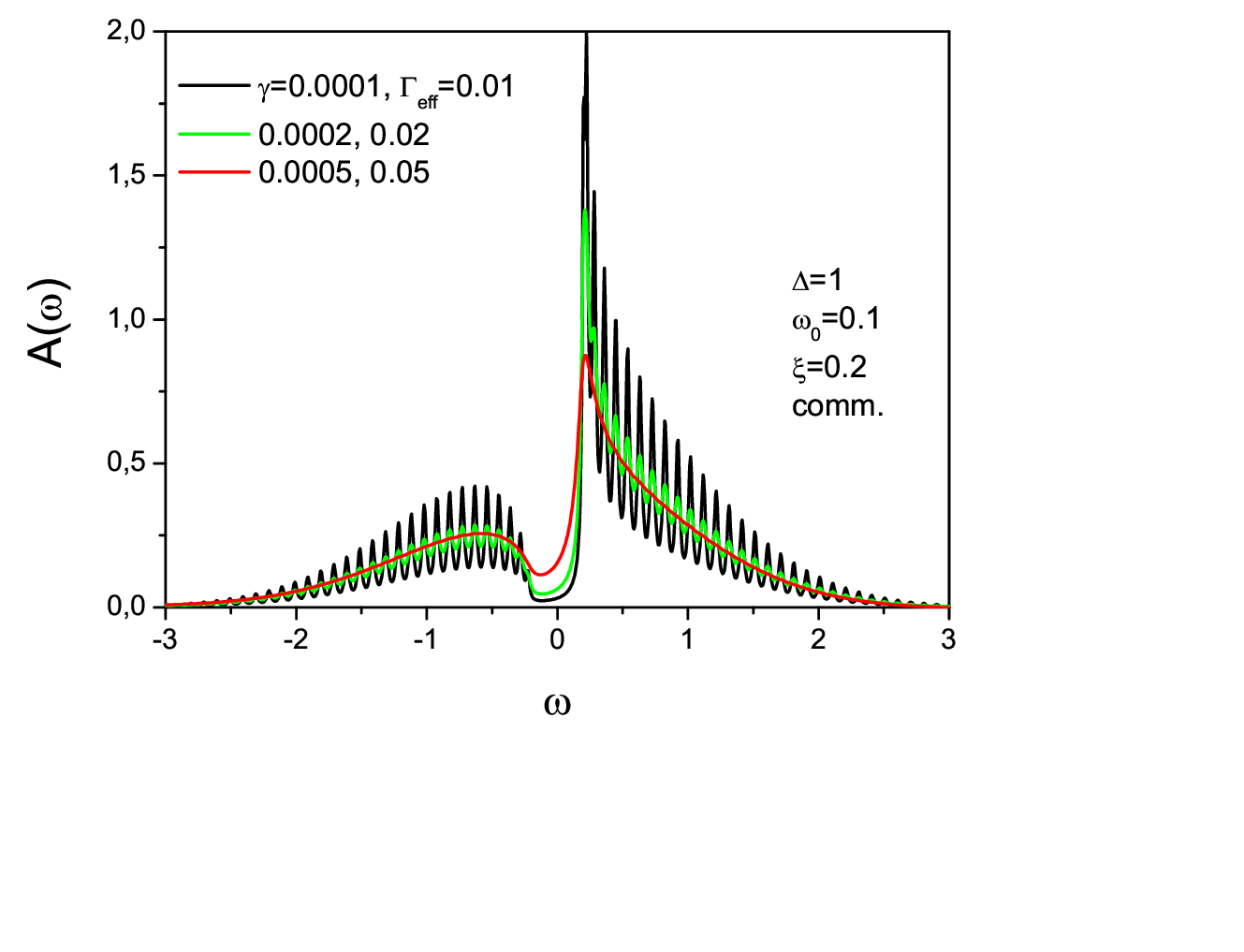}
\caption{Spectral density in commensurate case and $\omega_0=0.1$
for different values of electron energy $\xi$. The last panel shows
$\gamma$ dependence of spectral density at $\xi=0.2$. Black lines on first
three panels represent the results for spectral density obtained for static 
model ($\omega_0=0$) from exact continued fractions. All energies are given 
in units of $\Delta$.}
\label{SPD_comm_1}
\end{figure*}

\section{Exact Continued-Fraction Solution in the Quasistatic Regime}

As discussed in the previous sections, the coherent double-series
representation provides a natural description of the dynamic pseudogap
problem whenever the characteristic oscillation frequency of the order
parameter exceeds the dynamically generated decoherence scale,
\begin{equation}
\omega_0>\Gamma_{\rm eff}.
\end{equation}
In the opposite regime,
\begin{equation}
\Gamma_{\rm eff}\gtrsim\omega_0,
\label{QuasiStaticCriterion}
\end{equation}
the electron loses phase coherence before it can resolve the oscillatory
motion of the fluctuating field. The dynamics therefore becomes
effectively quasistatic, and the finite frequency $\omega_0$ may be
neglected in the correlation function. The problem then reduces to a
dynamic model with exponentially decaying temporal correlations,
\begin{equation}
\left<
\Delta(t)\Delta^*(0)
\right>
=
\Delta^2
e^{-\gamma |t|},
\end{equation}
which admits an exact diagrammatic solution.

It is important to emphasize that this continued-fraction solution should
not be regarded as an independent approximation. Rather, it represents
the decohered infrared limit of the general dynamic pseudogap problem,
complementary to the coherent double-series representation discussed
above.

\subsection{Exact diagrammatic summation}

The derivation follows the method originally developed for the
one-dimensional static pseudogap problem by one of the present authors
\cite{Sad3,SadovskiiBook}. The essential simplification of the present
dynamic model is that all internal frequency integrations can be
performed exactly by contour integration.

Indeed, every interaction line contributes a Lorentzian frequency
dependence possessing only simple poles in the complex frequency plane.
As a consequence, every internal frequency integration is evaluated
exactly by residues. This should be contrasted with the original static
one-dimensional model, where analogous momentum integrations required an
asymptotically controlled approximation \cite{SK98}. In this sense the
present dynamic model is exactly solvable.

A remarkable property of the model is that every diagram containing
crossing interaction lines is exactly equal to a noncrossing diagram
obtained by a suitable relabeling of interaction vertices
\cite{Sad3,SadovskiiBook}. Consequently, the complete perturbation series
may be generated by considering only rainbow diagrams, provided every
interaction line is multiplied by an appropriate combinatorial factor
$v(k)$ counting the number of topologically equivalent diagrams.

For commensurate fluctuations,
\begin{equation}
v(k)=k,
\label{vcomm}
\end{equation}
whereas for incommensurate fluctuations
\begin{equation}
v(k)=
\left\{
\begin{array}{ll}
\dfrac{k+1}{2},
&
k\ {\rm odd},
\\[2mm]
\dfrac{k}{2},
&
k\ {\rm even}.
\end{array}
\right.
\label{vinc}
\end{equation}
These factors incorporate exactly the contributions of all crossed
diagrams.

\subsection{Continued-fraction recursion}

Summation of the complete perturbation series then reduces to the exact
recursion relation for the irreducible self-energy,
\begin{equation}
\Sigma_k(\omega,\xi)
=
\frac{\Delta^2v(k)}
{\omega-(-1)^k\xi
+ik\gamma
-\Sigma_{k+1}(\omega,\xi)},
\label{Si_rec}
\end{equation}
which immediately yields the recursion relation for the Green's function,
\begin{equation}
G_k(\omega,\xi)
=
\left[
\omega
-
(-1)^k\xi
+
ik\gamma
-
\Delta^2v(k+1)
G_{k+1}
(\omega,\xi)
\right]^{-1}.
\label{G_rec}
\end{equation}
The physical Green's function is obtained from
$G(\omega,\xi)=G_0(\omega,\xi)$.
Equivalently, one obtains the exact continued-fraction representation
\begin{widetext}
\begin{equation}
G(\omega,\xi)
=
\frac{1}
{\displaystyle
\omega-\xi-
\frac{v(1)\Delta^2}
{\displaystyle
\omega+\xi+i\gamma-
\frac{v(2)\Delta^2}
{\displaystyle
\omega-\xi+2i\gamma-
\frac{v(3)\Delta^2}
{\displaystyle
\omega+\xi+3i\gamma-\cdots}}}}.
\label{G_chain}
\end{equation}
\end{widetext}
Equations (\ref{Si_rec})--(\ref{G_chain}) provide the exact solution of
the quasistatic dynamic pseudogap model.

\subsection{Relation to the coherent theory}

The present work therefore possesses two complementary nonperturbative
descriptions.
For
\begin{equation}
\omega_0>\Gamma_{\rm eff},
\end{equation}
the coherent double-series representation provides a controlled
description of dynamically resolved sidebands.
Conversely, when
\begin{equation}
\Gamma_{\rm eff}\gtrsim\omega_0,
\end{equation}
the continued-fraction solution becomes asymptotically exact because the
electron experiences the fluctuating order parameter as effectively
quasi static during its coherence time.

Together, the coherent double-series representation and the exact
continued fraction constitute complementary asymptotic descriptions of
the same underlying dynamic pseudogap model. As will be demonstrated
below, the crossover between the two regimes is accurately controlled by
the ratio $\omega_0/\Gamma_{\rm eff}$.

\begin{figure*}
\includegraphics[clip=true,width=0.45\textwidth]{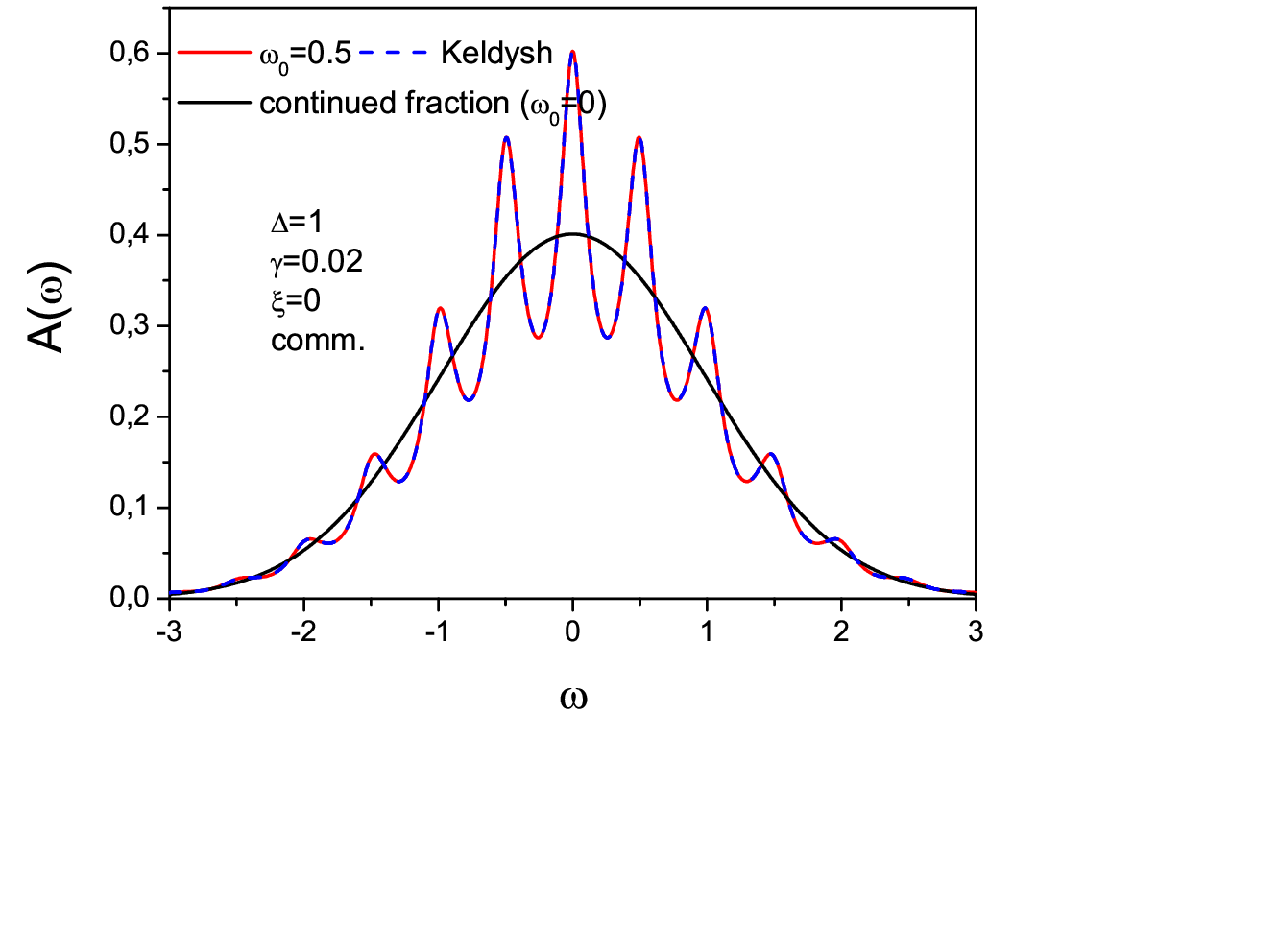}
\includegraphics[clip=true,width=0.45\textwidth]{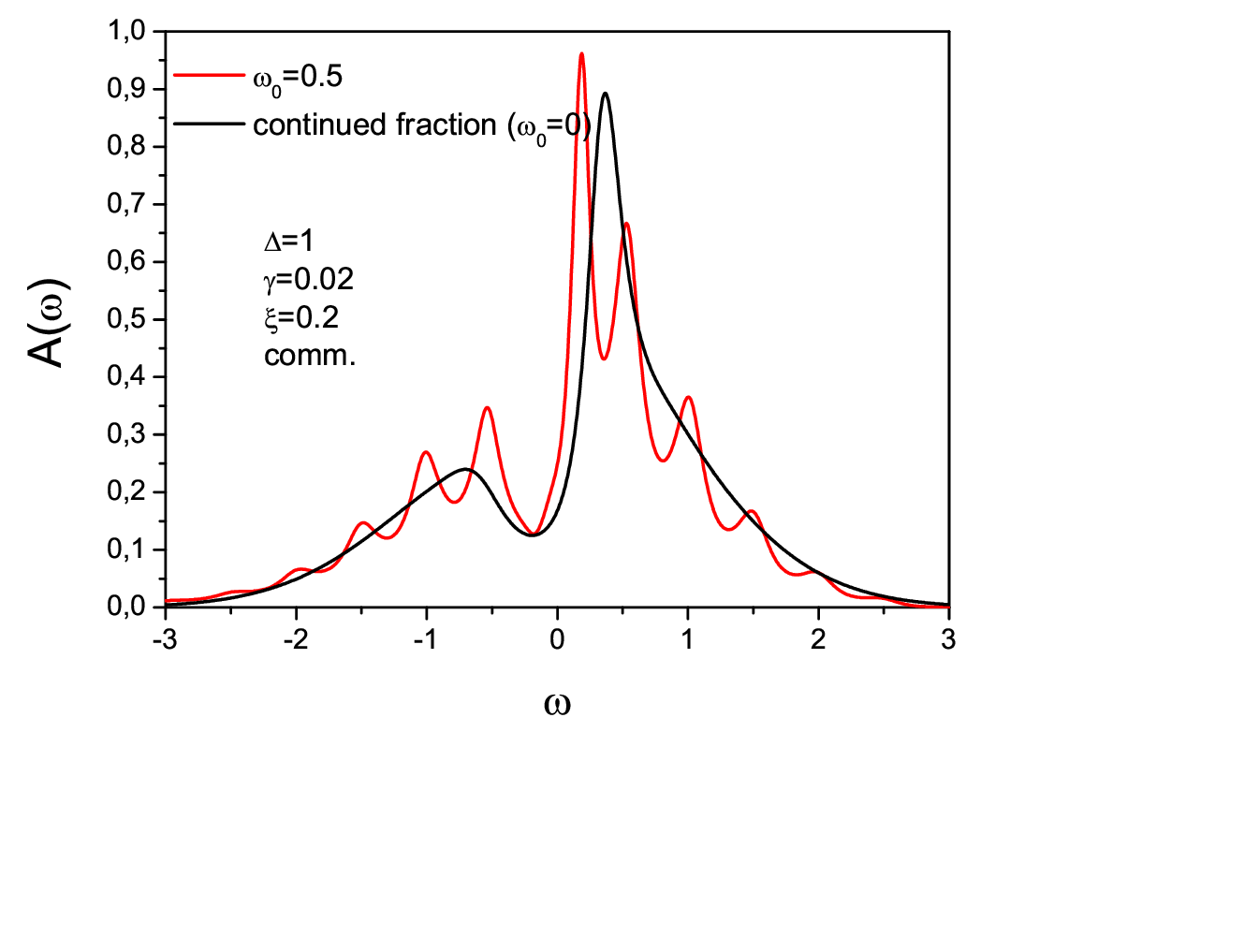}
\includegraphics[clip=true,width=0.45\textwidth]{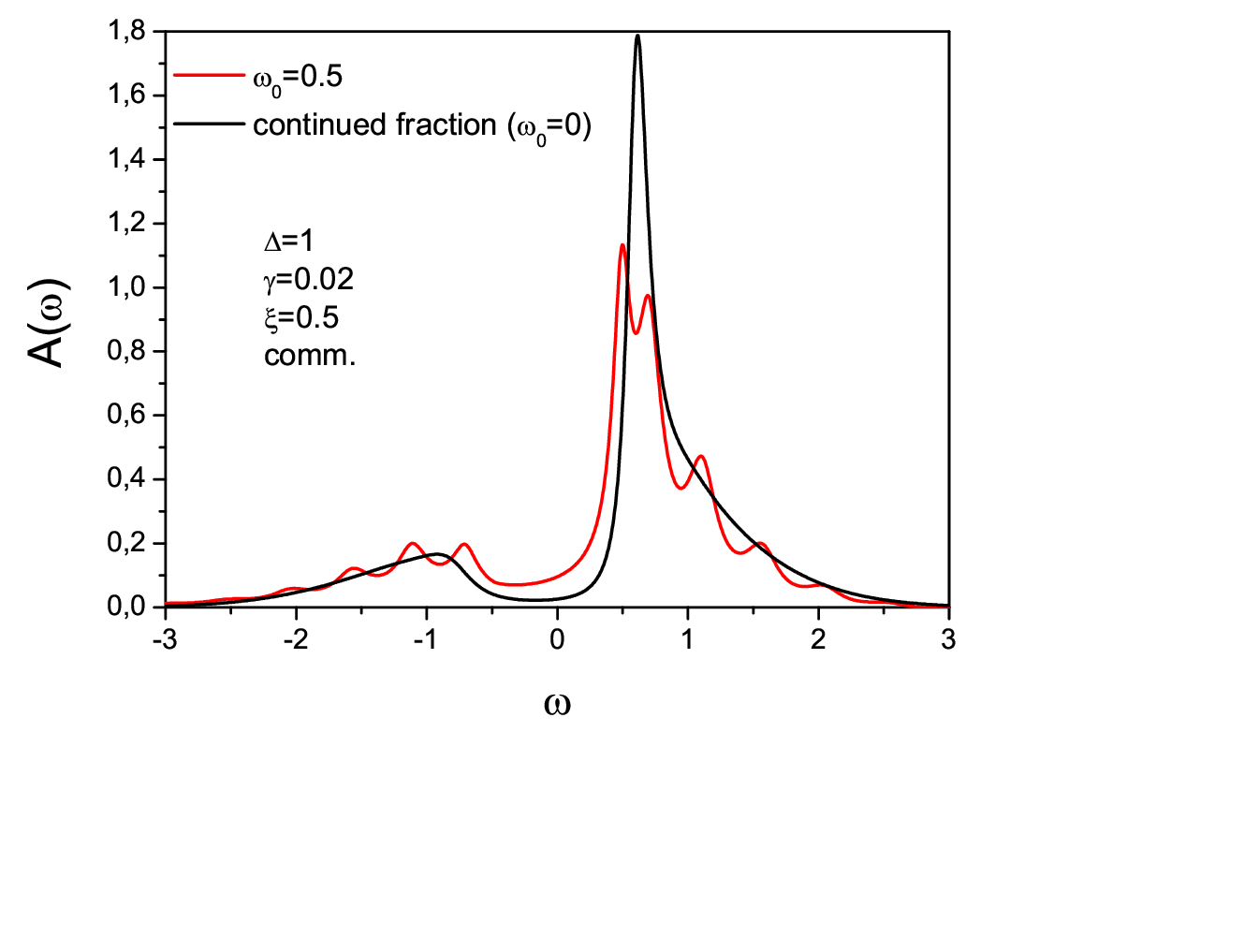}
\includegraphics[clip=true,width=0.45\textwidth]{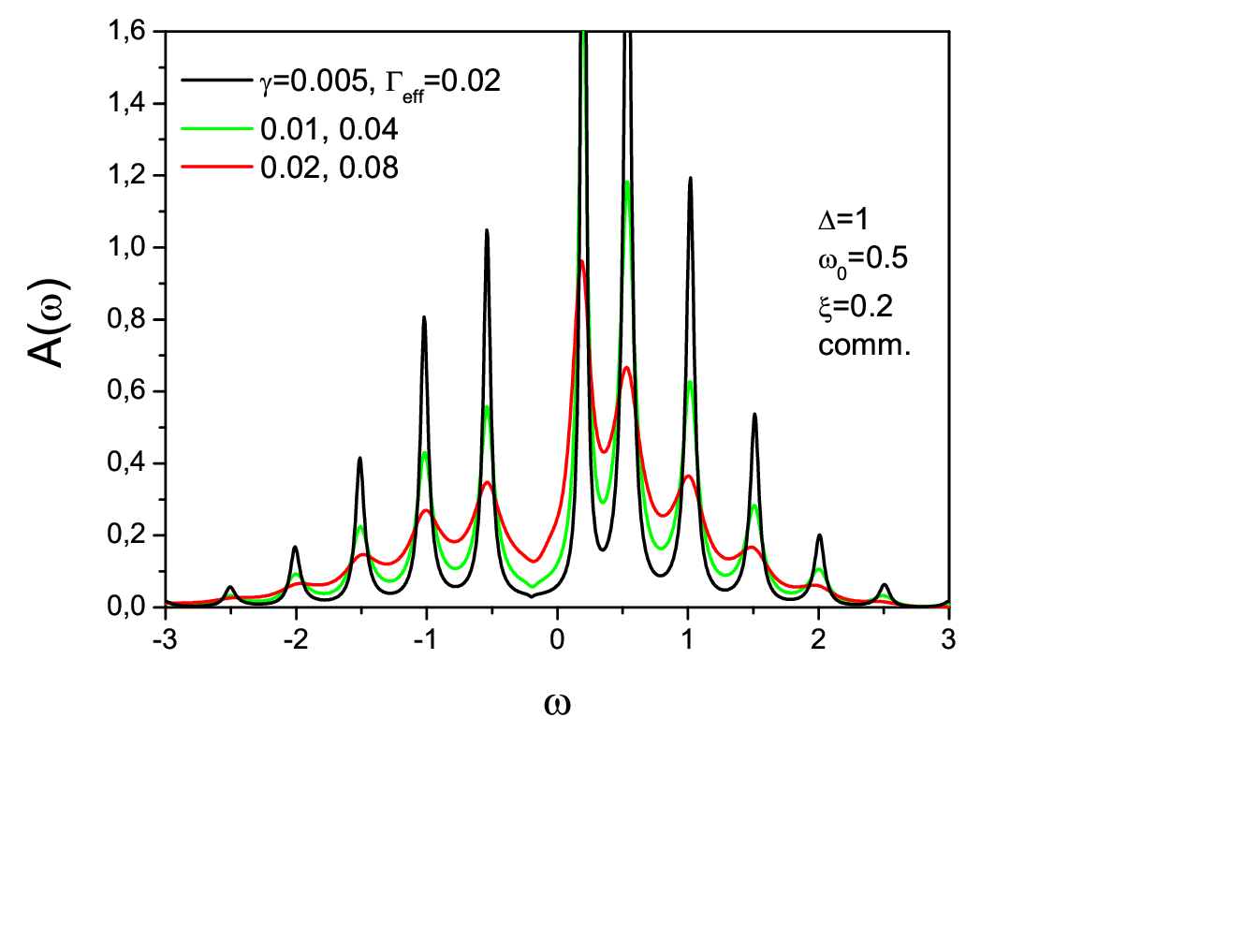}
\caption{Spectral density in commensurate case of dynamic pseudogap model
with $\omega_0=0.5$ for different values of electron energy $\xi$. 
Black lines on first three panels represent the exact results for spectral 
density obtained for static model ($\omega_0=0$) from exact continued fractions. 
The dashed blue line in the first panel represents the exact solution for
Keldysh model ($Q=0$) \cite{KS2024}, while the red line shows the exact solution 
of dynamic pseudogap model for $\xi=0$. 
The last panel shows $\gamma$ dependence of 
spectral density for fixed $\xi=0.2$. 
All energies are given in units of $\Delta$.}
\label{SPD_comm_2}
\end{figure*}

\begin{figure*}
\includegraphics[clip=true,width=0.45\textwidth]{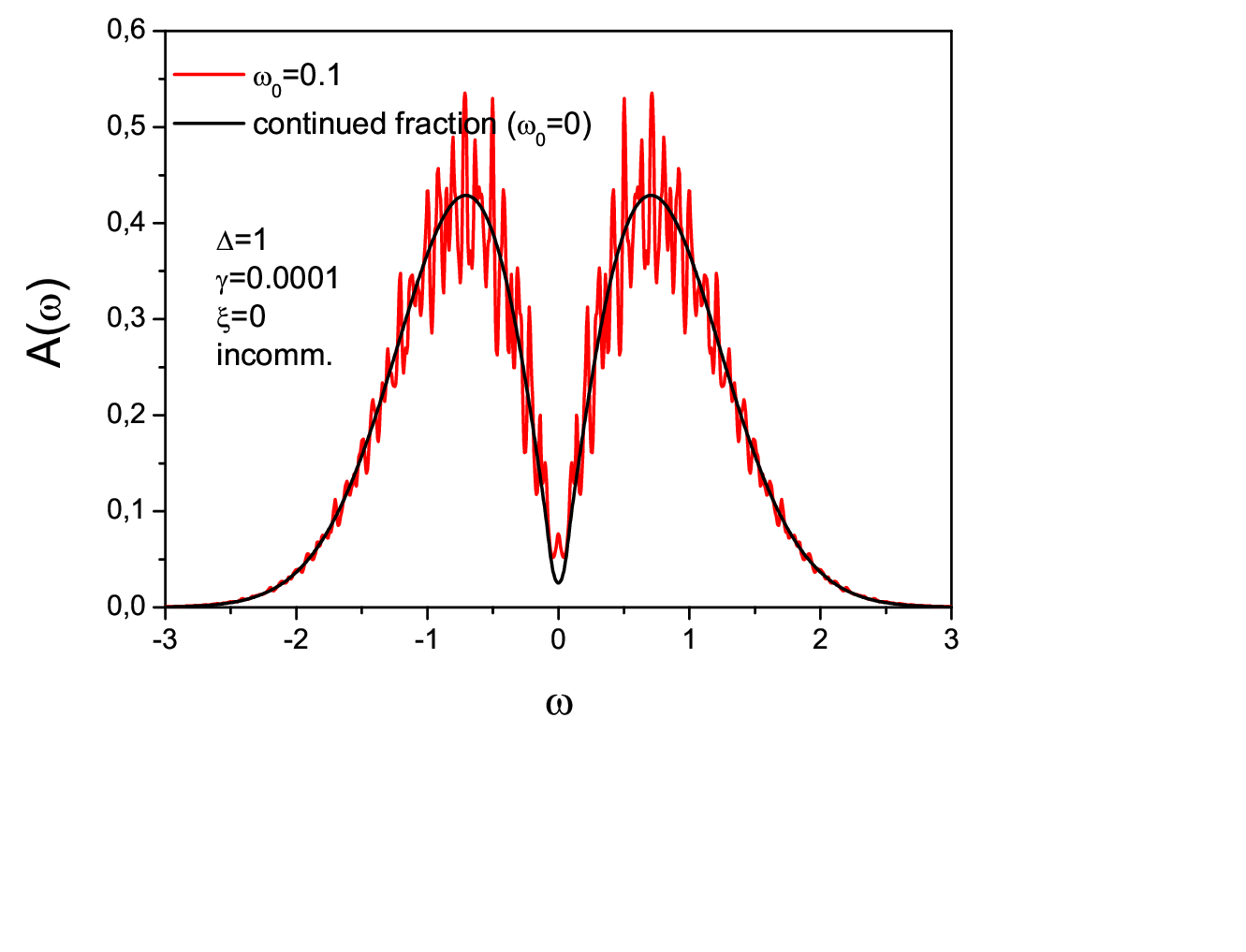}
\includegraphics[clip=true,width=0.45\textwidth]{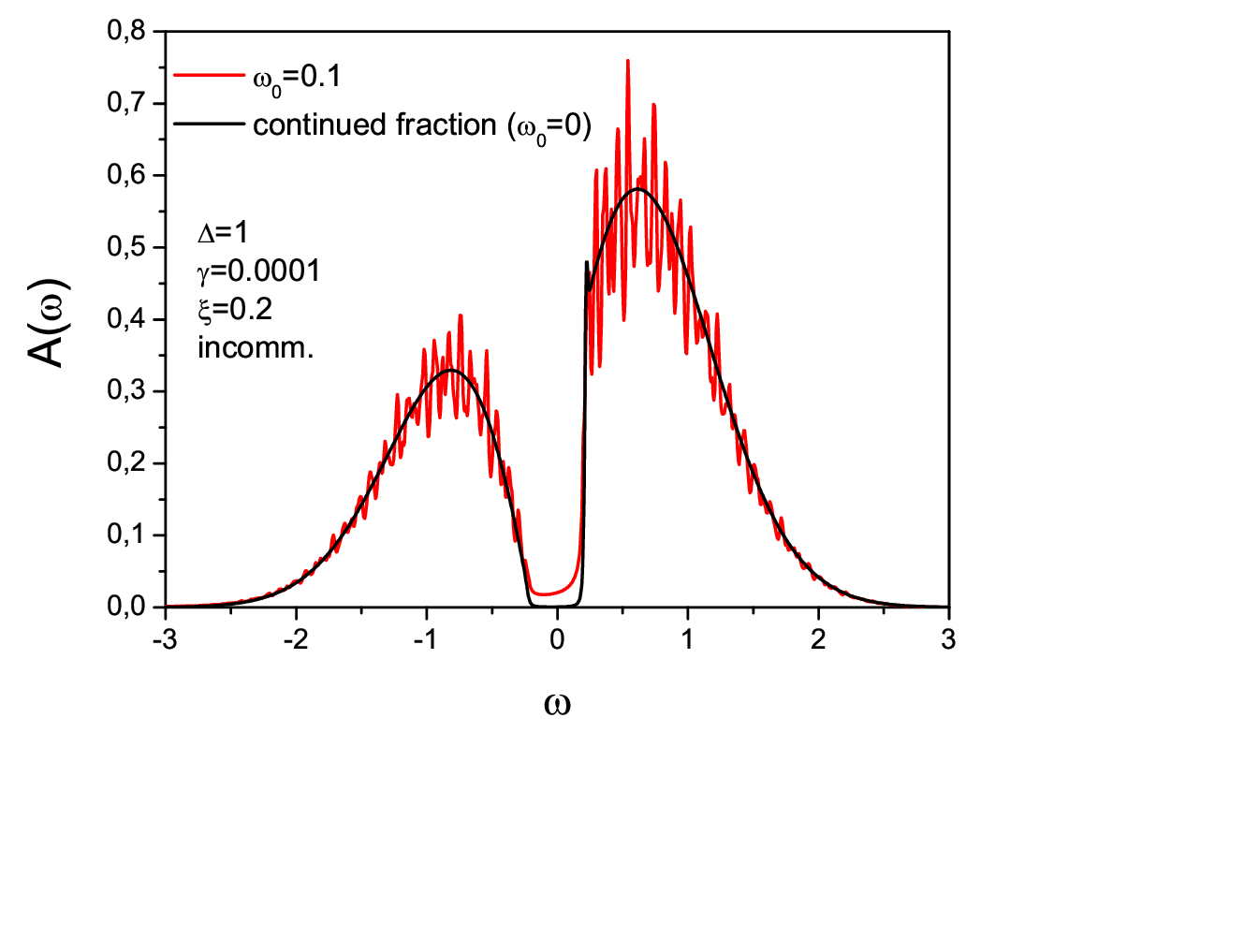}
\includegraphics[clip=true,width=0.45\textwidth]{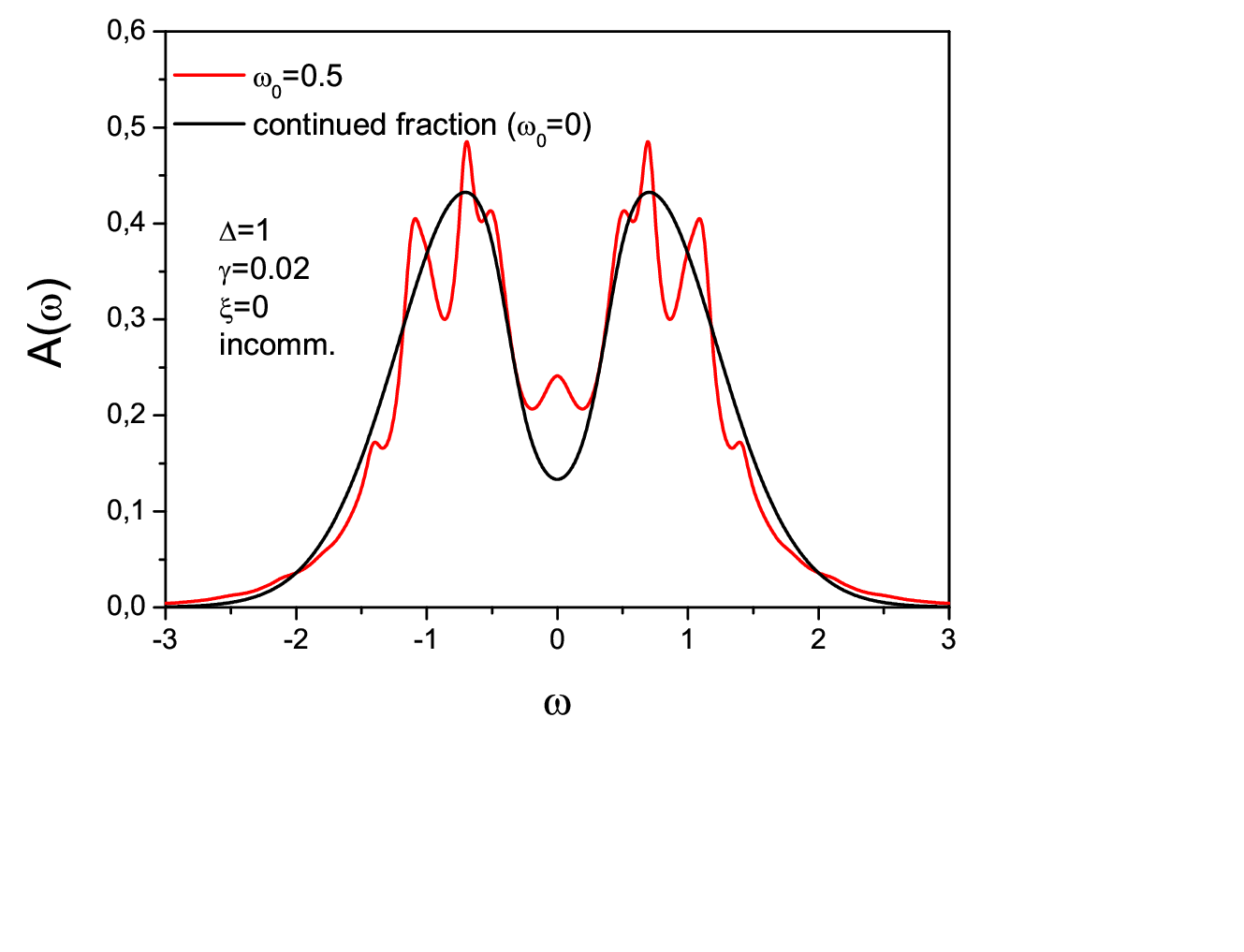}
\includegraphics[clip=true,width=0.45\textwidth]{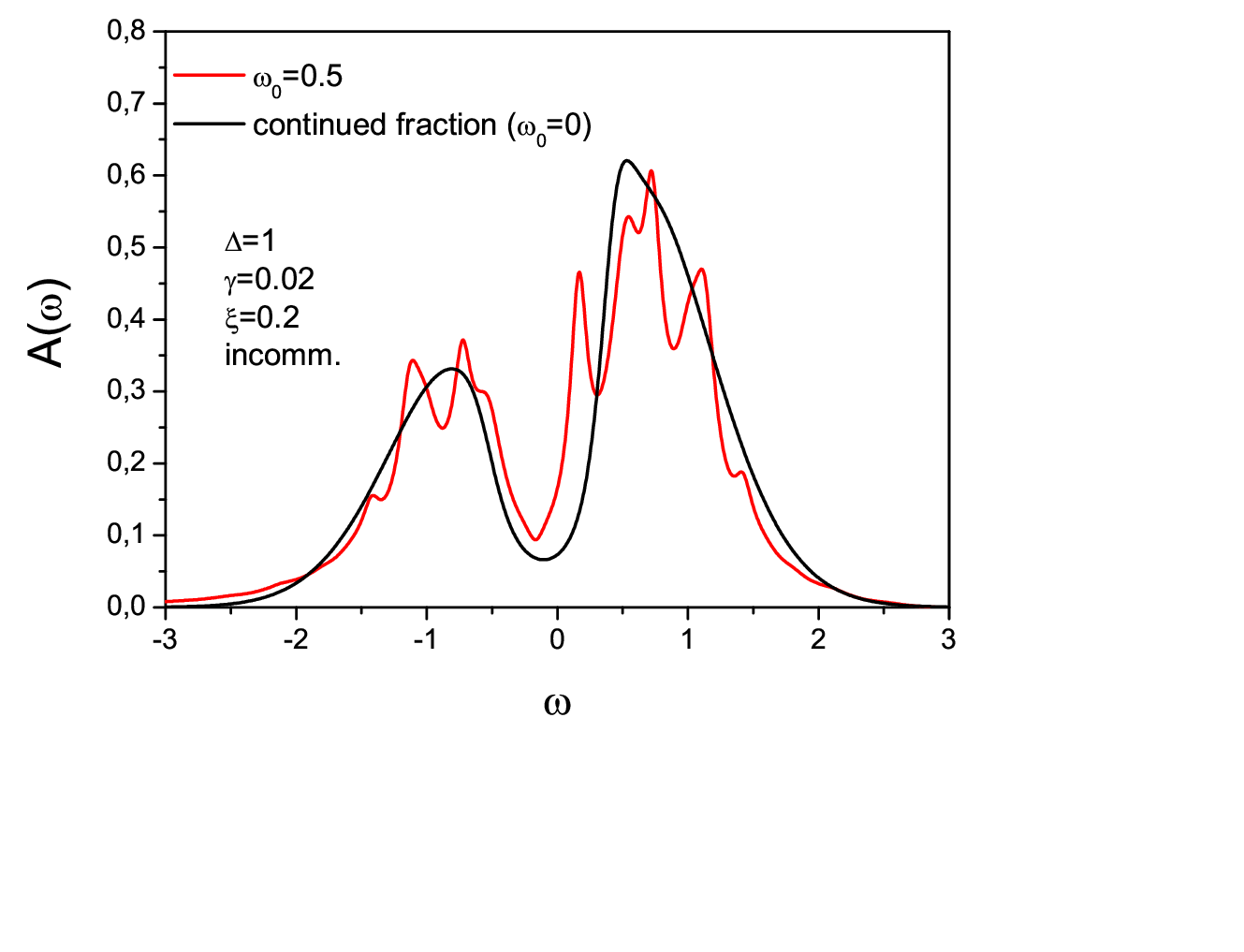}
\includegraphics[clip=true,width=0.45\textwidth]{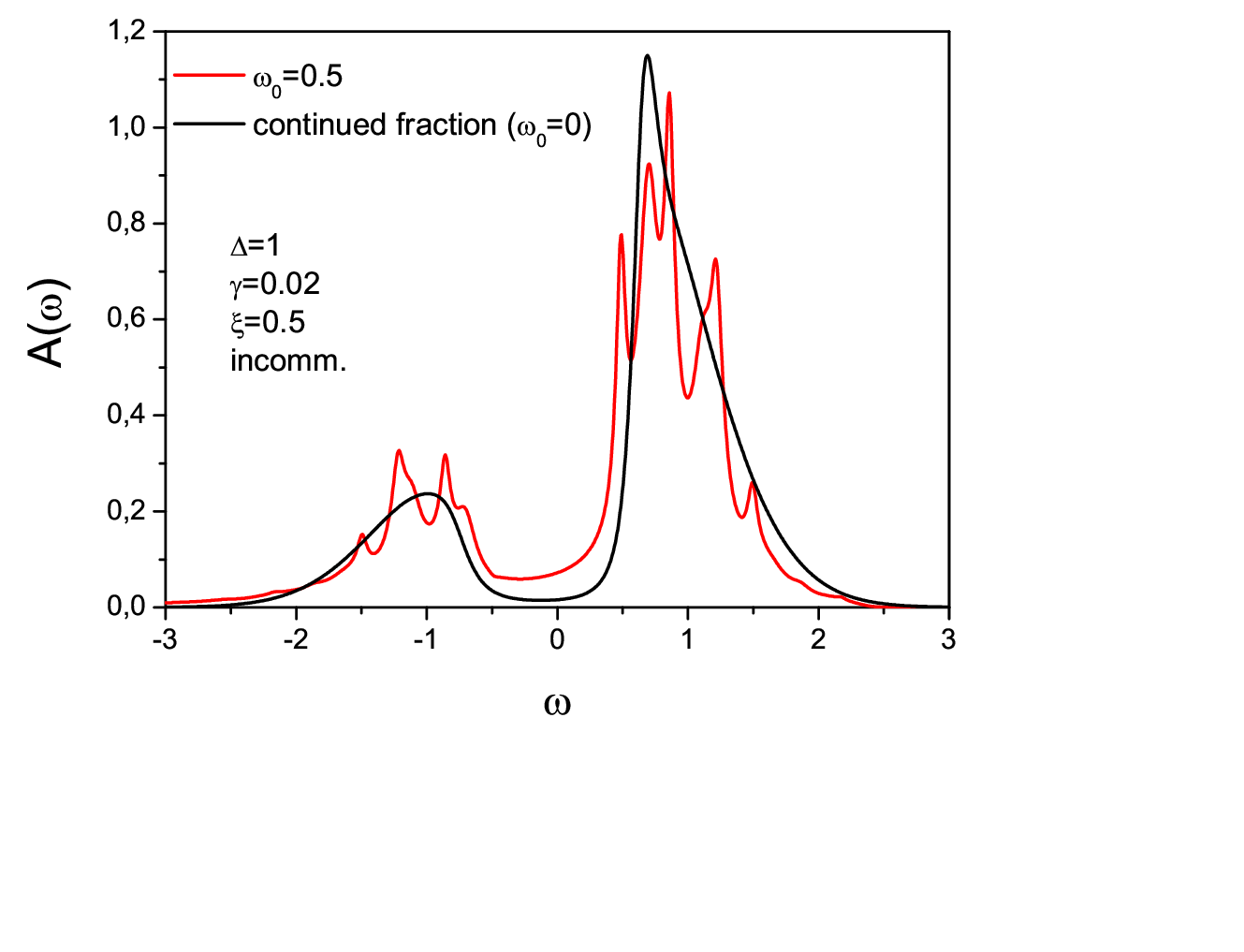}
\includegraphics[clip=true,width=0.45\textwidth]{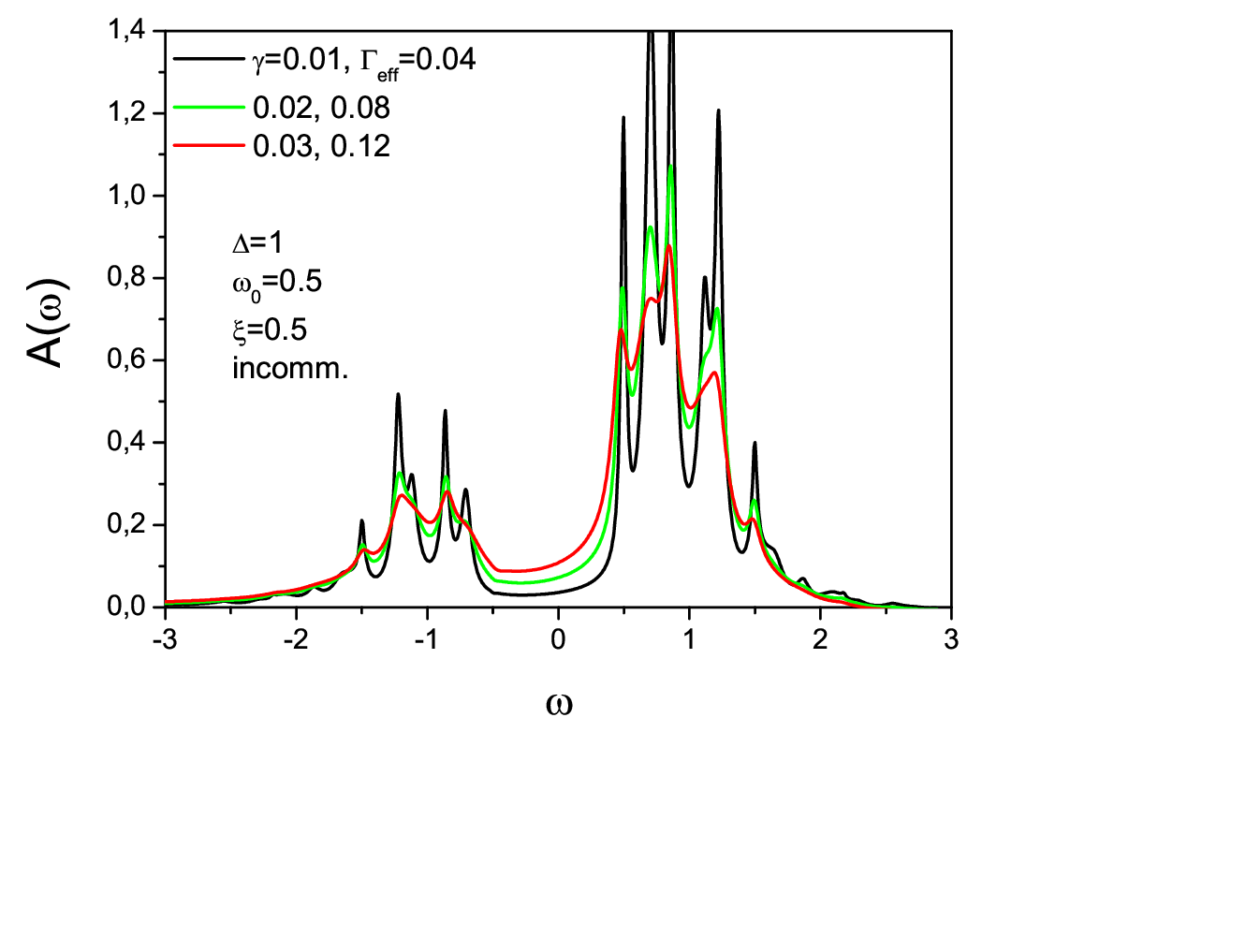}
\caption{Spectral density for incommensurate case and 
$\omega_0=0.1, 0.5$ for different values of electron energy $\xi$. 
Black lines represent the results for spectral density obtained for static model 
($\omega_0=0$) from exact continued fractions. 
Last panel shows dependence of spectral density on $\gamma$ at $\omega_0=0.5$ 
and $\xi=0.5$. All energies are given in units of $\Delta$.}
\label{SPD_incomm_1}
\end{figure*}

\begin{figure*}
\includegraphics[clip=true,width=0.45\textwidth]{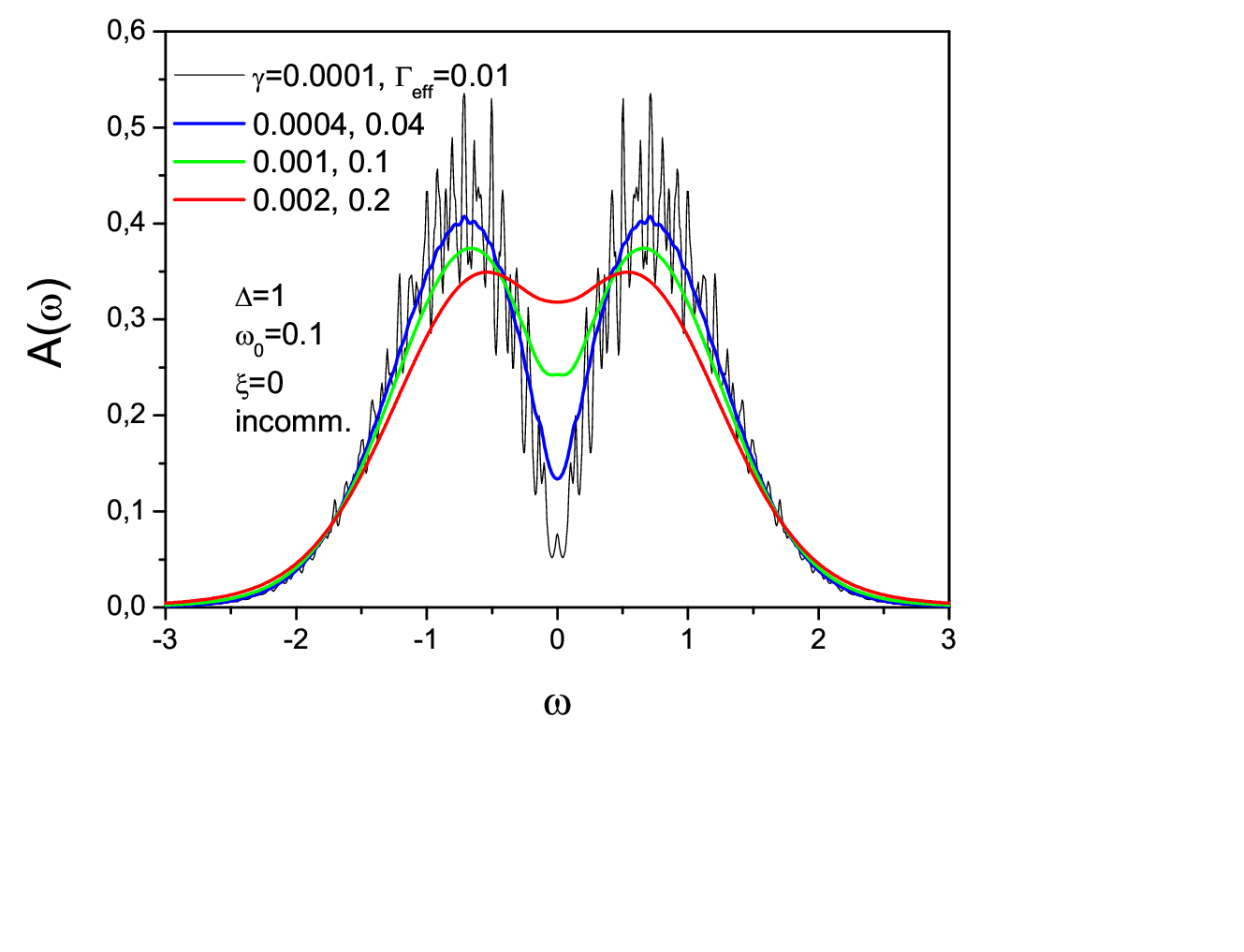}
\includegraphics[clip=true,width=0.45\textwidth]{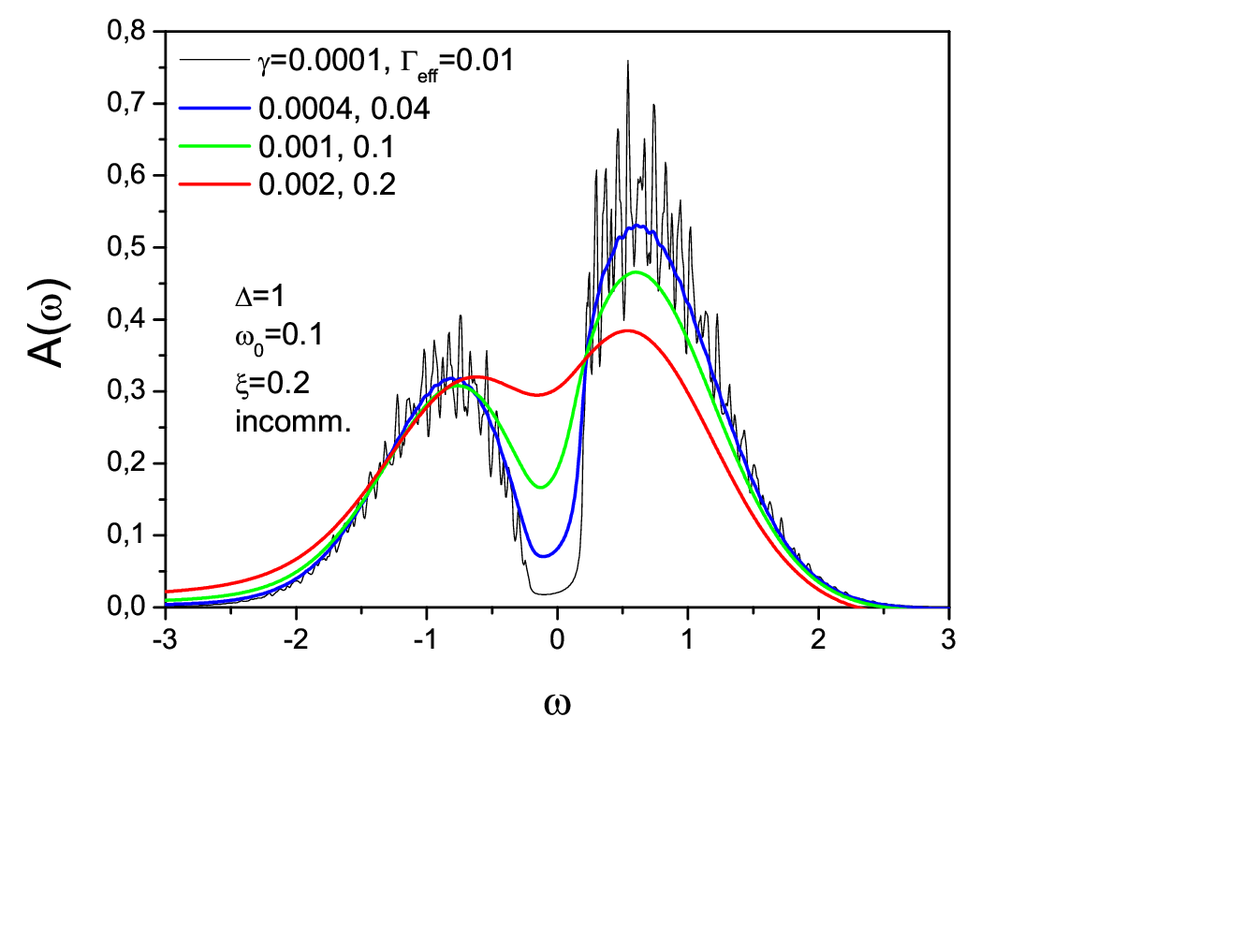}
\includegraphics[clip=true,width=0.45\textwidth]{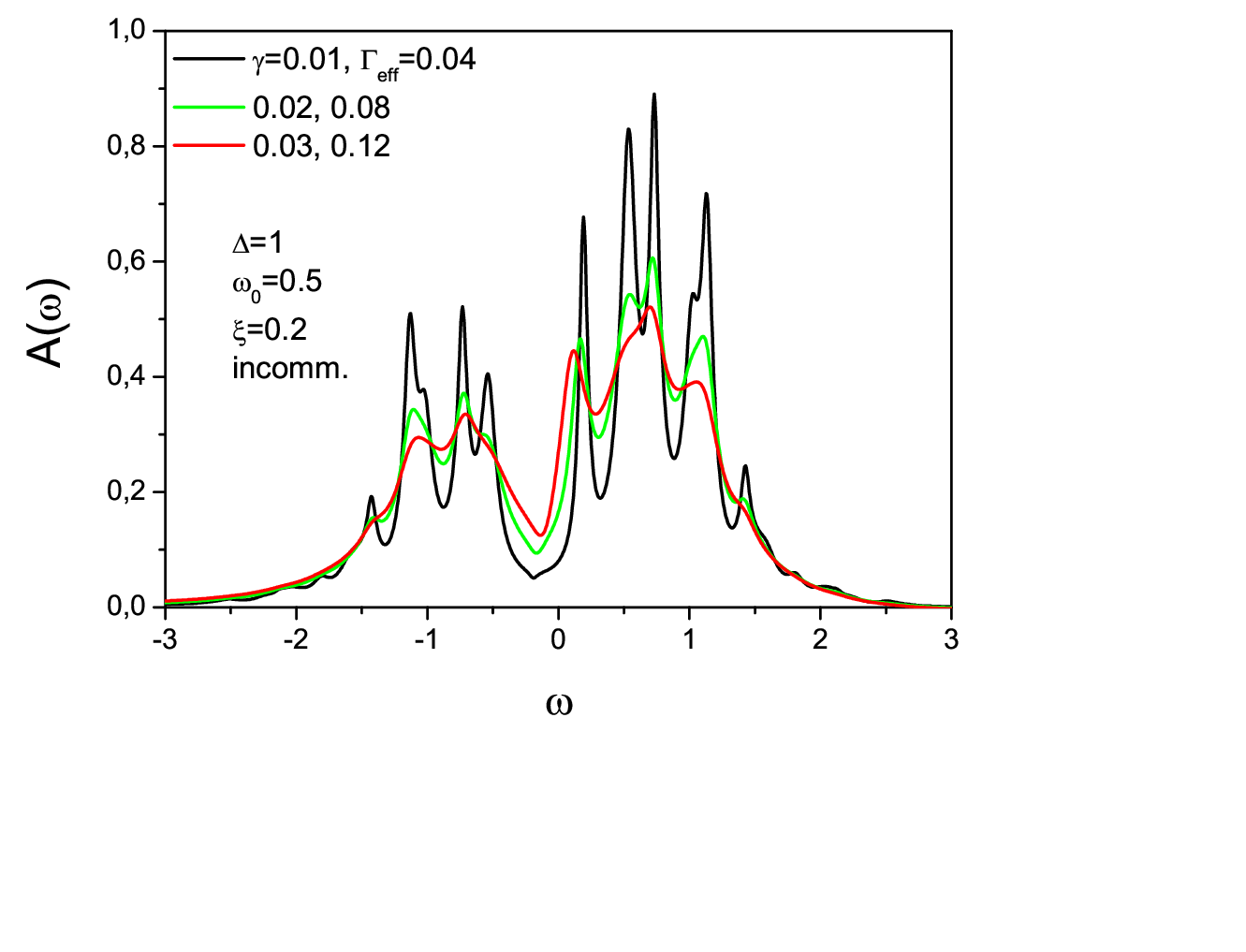}
\includegraphics[clip=true,width=0.45\textwidth]{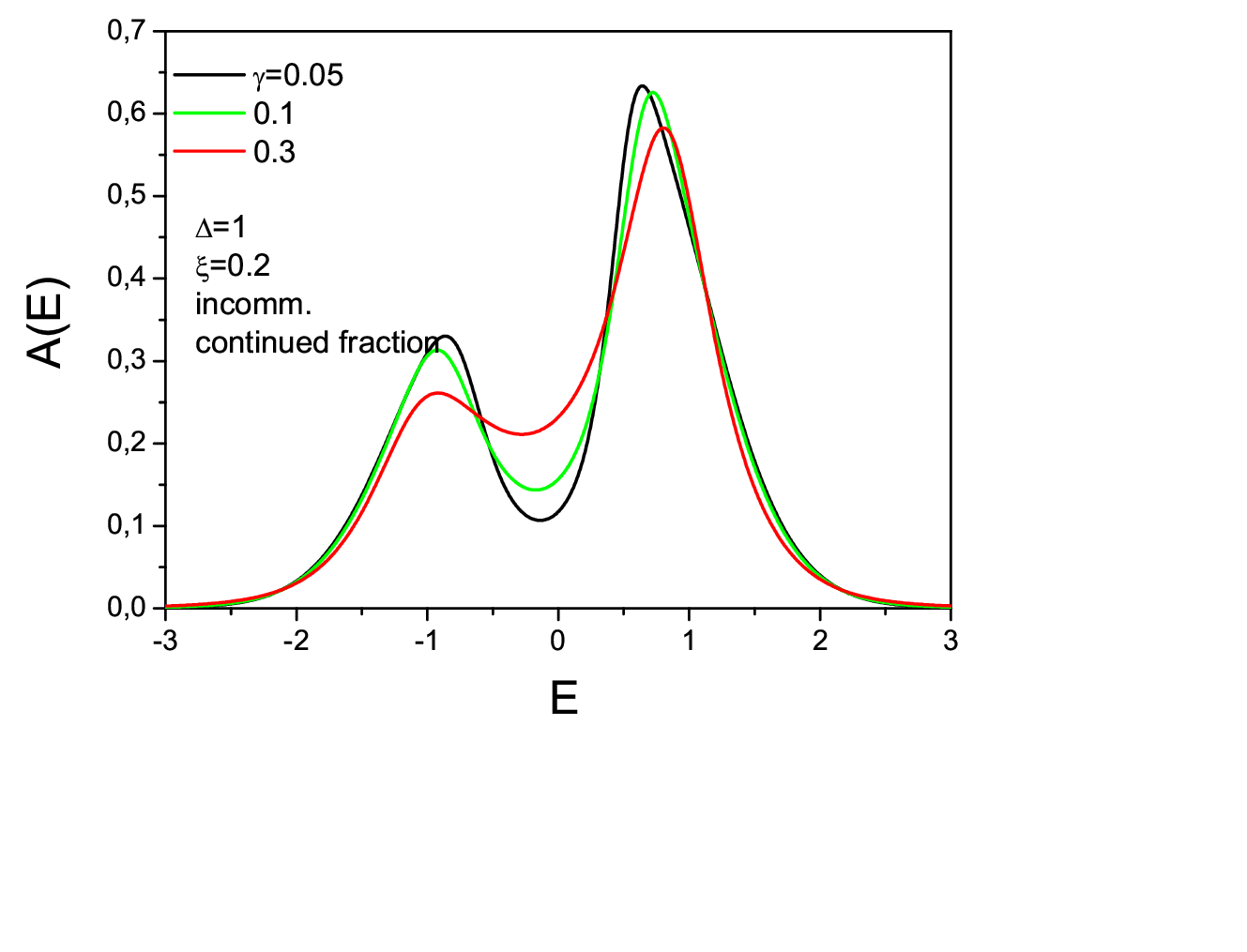}
\caption{Dependence of the spectral density for incommensurate case 
on $\gamma$ for different values of $\omega_0$ and $\xi$. 
The last panel
shows spectral density dependence on $\gamma$ in static ($\omega_0=0$)
approximation obtained from exact continued fraction representation.
All energies are given in units of $\Delta$.}
\label{SPD_incomm_2}
\end{figure*}

\section{Spectral Properties of the Dynamic Green's Function}

The double-series representation obtained above provides an explicit
analytical expression for the Green's function.
It is therefore important to analyze its fundamental analytical and
spectral properties.
The electronic spectral density is defined in the usual manner,
\begin{equation}
A(\omega,\xi)
=
-\frac{1}{\pi}
{\rm Im}
\,G^R(\omega,\xi).
\label{SpectralFunction}
\end{equation}
The general spectral sum rule is satisfied
\begin{equation}
\int_{-\infty}^{\infty}
A(\omega,\xi)
\,d\omega
=
1.
\label{SumRule}
\end{equation}
expressing conservation of the total electronic spectral weight.
Any numerical implementation of the double-series expansion should satisfy
this relation with high accuracy and may therefore use it as a convenient
convergence test.

The observable spectrum is obtained as a coherent superposition of
the spectral functions associated with individual dynamic histories.
For an exact retarded Green's function the spectral density satisfies
\begin{equation}
A(\omega,\xi)\ge0.
\end{equation}
However, this positivity property need not hold separately for individual
dynamic histories. Indeed, in our case:
\begin{itemize}
\item
the history weights are generally complex,
\item
the effective dynamic gaps are complex,
\item
the generalized Bogoliubov coherence factors also become complex.
\end{itemize}
Therefore individual terms of the double series should be interpreted as
coherent amplitudes rather than independent spectral contributions.
Only the complete resummed Green's function is expected to possess a
strictly positive spectral density.

\subsection{Numerical Results for Spectral Densities and Density of States}

We performed extensive numerical calculations of spectral functions of
our model (\ref{SpectralFunction}) using double series
representation for the Green's function (\ref{GorFreqDoubleSer}), 
(\ref{BogolFreqDoubleSer}) for dynamic model,  making comparison with exact
continued fraction solution for the static ($\omega_0=0$) case  (\ref{G_rec}). 
We have explicitly checked that the exact sum rule (\ref{SumRule}) is satisfied
precisely for all values of the parameters of the model used in our 
calculations.

The double-series representation is especially convenient for numerical
calculations and continued fraction algorithm demonstrates very fast convergence. 
Since the Poisson weights in double series decrease factorially, the sums may 
be truncated at finite values $n\le n_{\rm max}$, $m\le m_{\rm max}$,
with rapidly controllable accuracy. In our calculations we used
the values of $n_{\rm max},\, m_{\rm max}$ with $n_{\rm max}+m_{\rm max}\leq$ 
10$^3$. 
The principal numerical limitation is not convergence of the
series itself but rather the increasing cancellation between neighboring
histories in the strongly overdamped regime.
As discussed in the previous section, this behavior signals the crossover
from coherent dynamic sidebands toward the quasistatic fluctuating-gap
description.
Below we show only most typical results obtained for characteristic values
of the model parameters.

In Fig. \ref{SPD_comm_1} we show spectral densities for the case of 
commensurate fluctuations with relatively small values of the frequency
$\omega_0=0.1$ (all energies below are measured in units of the amplitude 
$\Delta$). First panel shows results at the Fermi surface $\xi=0$, where
our commuting solution is exact. Here we take very small values of damping 
$\gamma=0.0001$.
The next two panels show spectral density for $\xi\neq 0$, demonstrating 
typical asymmetric pseudogap behavior. In all these cases we observe 
characteristic modulation of spectral density with frequency $\omega_0$, 
around the smooth behavior obtained from exact continued fraction solution
for the case of $\omega_0=0$ and also shown in these graphs. 
The last panel of Fig. \ref{SPD_comm_1} shows $\gamma$ dependence of spectral 
density and corresponding values of $\Gamma_{\mathrm{eff}}$, demonstrating 
smooth suppression of sidebands (modulations). Actually these results are very 
similar to those obtained for spectral density in $Q=0$ Keldysh model 
in Ref. \cite{KS2024}.
Analogous results for larger $\omega_0=0.5$ and $\gamma=0.02$ are shown in 
Fig. \ref{SPD_comm_2}, where at the last panel we again show the $\gamma$
dependence and suppression of modulations at larger damping. 

We have specifically checked that in the limit of large $\gamma$ our results
for $\xi=0$ smoothly transform to an exact static spectral density calculated
from continued fraction solution, confirming the exactness of double-series
solution in this case. 

In case of the finite $\xi$, 
good correlation with continued fraction results is observed only for 
$\omega_0\gg\Gamma_{\mathrm{eff}}$. The deviations appearing for large enough
$\gamma$ when $\omega_0\sim\Gamma_{\mathrm{eff}}$ are attributed to our 
neglect of the Magnus terms contributions in our commuting approximation. 
However, it is obvious on physical grounds that for large values of 
$\Gamma_{\mathrm{eff}}>\omega_0$ the exact static continued fraction solution 
gives reliable description of spectral density for our dynamic model.

Note that the double-series representation follows from the exact cumulant 
solution of the commuting sector of the dynamic finite-$Q$ problem.
Its validity therefore depends upon the extent to which the neglected
non-Abelian SU(2) corrections remain small.

The behavior of spectral densities in case of incommensurate fluctuations is
in some sense similar but more complicated.
In Fig. \ref{SPD_incomm_1} we show spectral densities for the case of 
incommensurate fluctuations for small values of the frequency
$\omega_0=0.1$. First two panels shows results for $\xi=0$ and $\xi=0.2$
and very small values of damping $\gamma=0.0001$. We observe typical pseudogap
behavior of spectral density but with seemingly chaotic modulations with
frequency $\omega_0$. The next two panels show spectral densities at larger
values of $\gamma=0.02$. Actually in all cases here modulation is regular
but peaks at $\omega_0$ harmonics are splitted due to the existence of two
independently fluctuating fields for incommensurate case, which creates more
complicated picture of spectral density modulations as compared with
commensurate case. 

In Fig. \ref{SPD_incomm_2} we show the $\gamma$ dependence of spectral density 
for incommensurate case. 
In general we observe the suppression of sidebands 
in the spectral density with growth of $\gamma$, similar to commensurate case.

Results obtained from double series are again reliable for
$\omega_0\gg\Gamma_{\mathrm{eff}}$ and continued fraction description is
valid for $\Gamma_{\mathrm{eff}}>\omega_0$, with results for this region
shown on the last panel of Fig. \ref{SPD_incomm_2}, while the crossover region
of $\omega_0\sim\Gamma_{\mathrm{eff}}$ can not be reliably 
described due to the role of neglected Magnus terms. However, it is absolutely 
clear physically that this crossover is smooth and continuous.

\begin{figure*}
\includegraphics[clip=true,width=0.45\textwidth]{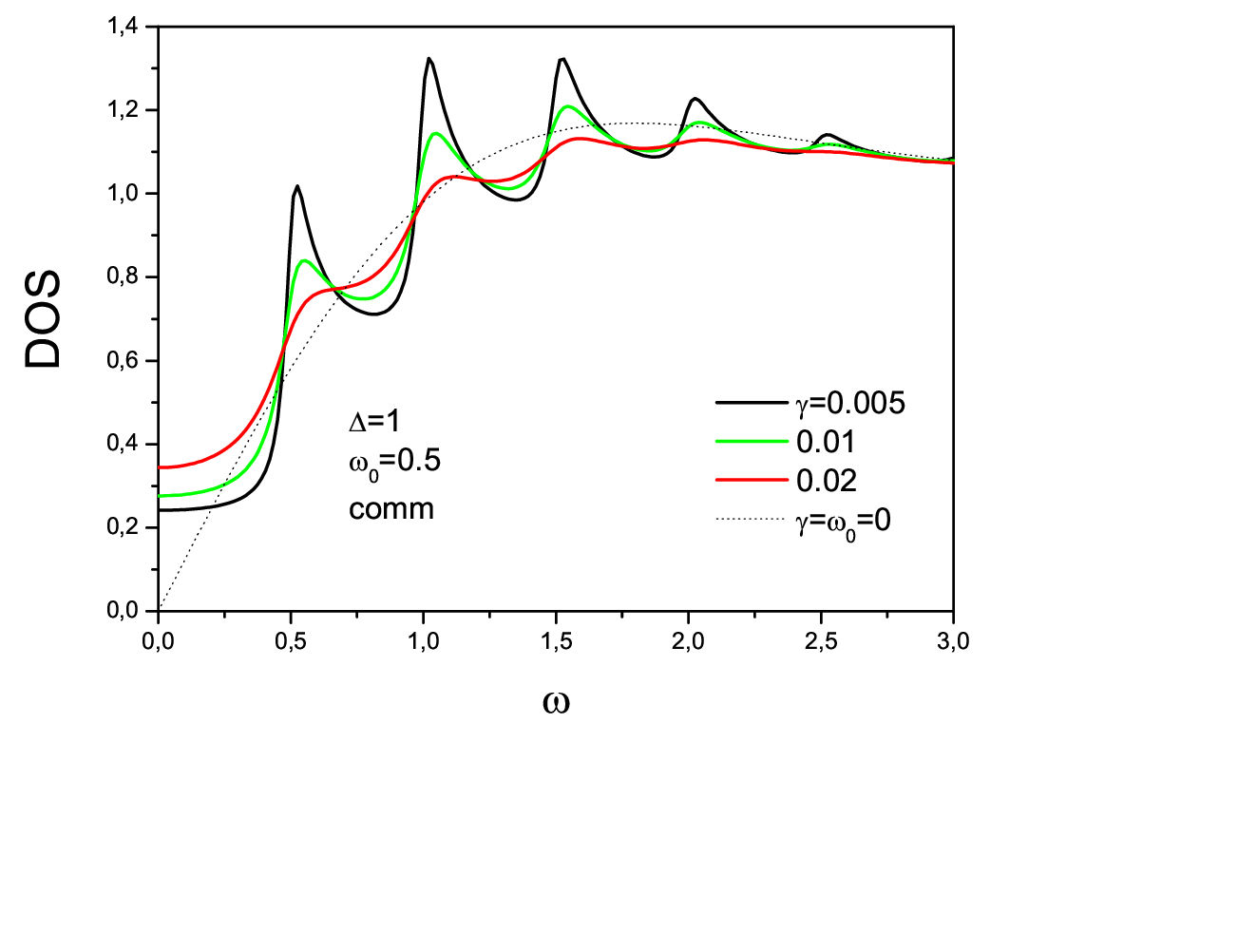}
\includegraphics[clip=true,width=0.45\textwidth]{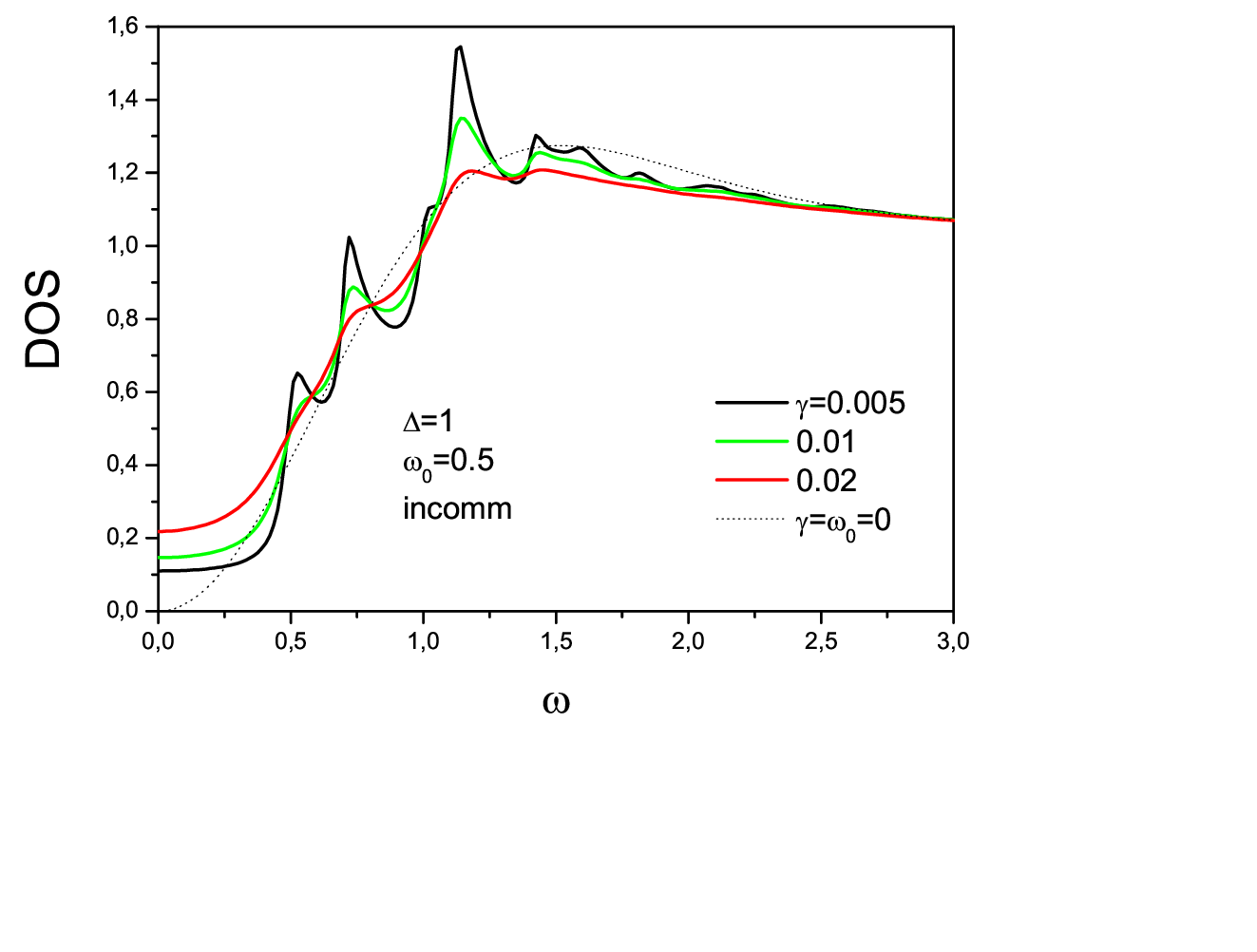}
\caption{Density of states (in units of free electron DOS) for for commensurate (left) 
and incommensurate (right) cases and typical values of $\omega_0$ and $\gamma$.
Dotted lines represent exact behavior for static case $\omega_0=0$ and
$\gamma=0$ (Appendix \ref{app:Static}).
}
\label{DOS_psgap}
\end{figure*}

Total  density of states can be directly calculated from its definition:
\begin{equation}
N(\omega)=-\frac{1}{\pi}\int_{-\infty}^{\infty}d\xi{\rm Im}\,G^R(\omega,\xi).
\label{DOS}
\end{equation}
For simplicity all calculations were performed for bare 
electronic band of infinite width. Exact nesting is assumed as usual in one --
dimensional case.  
Both for commensurate and incommensurate fluctuations the density of states 
demonstrates typical pseudogap behavior with additional peaks
due to sidebands (modulation of spectral density with frequency $\omega_0$)
as shown in Fig. \ref{DOS_psgap}, similar to that observed for Keldysh
$Q=0$ model in Ref. \cite{KS2024}. These peaks are gradually suppressed with the
growth of $\gamma$ as the pseudogap itself which closes at large enough
values of $\gamma$, as is well known for the case of static fluctuations
\cite{Sad3,SadTim,SK98}. This behavior is physically pretty obvious within
our approach.

Such additional peaks of the density of states are observable in principle
in tunneling experiments.

\subsection{General Physical Picture}

Thus our numerical results fully confirm the existence of the following
three physical regimes in our problem:

\subsubsection{Coherent Dynamic Regime}

The most favorable situation corresponds to relatively slow decoherence,
\begin{equation}
\omega_0
\gg
\Gamma_{\rm eff},
\label{CoherentCriterion}
\end{equation}
where $\Gamma_{\rm eff}$ denotes the dynamically generated linewidth.
In this regime neighboring sidebands remain well resolved,
for the dominant scattering histories.
The electron therefore retains phase coherence over many oscillation
periods of the fluctuating field.
The physical spectrum is naturally interpreted as a coherent superposition
of dynamically broadened Bogoliubov quasiparticles, and numerical
calculations indicate rapid convergence of the double-series expansion.

\subsubsection{Crossover Regime}

This is the most complicated region where
\begin{equation}
\Gamma_{\rm eff}
\sim
\omega_0.
\label{CrossCriterion}
\end{equation}
Neighboring sidebands begin to overlap and can no longer be regarded as
independent coherent excitations. In this region our double series 
in general overestimates the role of $\gamma$ and can lead to unphysical
results. Apparently, to obtain the complete description of crossover region 
we have to take into account non-commuting Magnus contributions.

\subsubsection{Overdamped Regime}

The situation changes qualitatively when
\begin{equation}
\Gamma_{\rm eff}\gg
\omega_0.
\label{OverdampedCriterion}
\end{equation}
Neighboring sidebands overlap strongly and can no longer be regarded as
independent coherent excitations.
Although the mathematical series itself continues to converge because of
the factorial suppression of the history weights, the physical
interpretation in terms of coherent scattering histories becomes
progressively less appropriate. 
However, in this case we can safely neglect $\omega_0$ and use the
continued-fraction solution for the static case instead, which correctly
reproduces the spectral density behavior in the more or less trivial limit 
of large $\gamma$. Actually this is fully confirmed by the numerical
results following from our double series expansion for the Green's function at
large $\gamma$.

Thus the numerical results obtained in the present work suggest the
following physical picture.
When the fluctuating field oscillates sufficiently rapidly compared with
its intrinsic decoherence rate, the electron is able to distinguish the
individual dynamical scattering events.
The corresponding spectrum consists of coherent dynamic sidebands.
As decoherence increases, repeated scattering histories lose their mutual
phase coherence.
The electron no longer resolves the temporal structure of the fluctuating
field and instead experiences transformation to effectively quasistatic
($\omega_0=0$) random gap.

\section{Physical Realizations}

The dynamic pseudogap model developed in the present work is intentionally
formulated in a phenomenological manner.
Its essential ingredients are a finite ordering wave vector ${\bf Q}$, a
temporally fluctuating classical order parameter, and sufficiently strong
electron scattering between nested portions of the Fermi surface.
These ingredients naturally occur in a wide variety of correlated electron
systems approaching density-wave instabilities.

\subsection{Charge-Density-Wave Fluctuations}

Perhaps the most direct realization of the present theory is provided by
electronic systems close to a Peierls or charge-density-wave (CDW)
transition.
Near the transition temperature the soft phonon mode associated with the
ordering vector ${\bf  Q}$ becomes strongly populated,
$\omega_{\bf Q}\rightarrow0$, and its occupation number
\begin{equation}
n_B(\omega_{\bf Q})
\simeq
\frac{T}{\omega_{\bf Q}}
\end{equation}
becomes very large.
Under these conditions the lattice displacement may be treated as a
classical stochastic field with finite temporal correlations
$\Delta(t)=g_{\bf Q} u_{\bf Q}(t)$,
where $u_{\bf Q}(t)$ denotes the slowly fluctuating soft phonon coordinate.

The phenomenological correlation function adopted throughout the present
work,
\begin{equation}
\langle
\Delta(t)\Delta^*(0)
\rangle
=
\Delta^2
e^{-\gamma|t|}
\cos{(\omega_0t)},
\end{equation}
may therefore be viewed as the effective long-time dynamics of a damped
soft phonon mode.

More estimates of this kind are described in Appendix \ref{app:Microscopic}.

\subsection{Spin-Density-Wave Fluctuations}

An analogous situation occurs near antiferromagnetic or spin-density-wave
(SDW) instabilities \cite{PS1,PS2,Chub_1,Chub_2,Mor_85}.
 In this case the fluctuating order parameter represents slowly varying
staggered magnetization,
\begin{equation}
\mathbf{M}(\mathbf{r},t)
=
\mathbf{M}_Q(t)
e^{i\mathbf{Q}\cdot\mathbf{r}}
+
\mathrm{c.c.},
\end{equation}
which again produces repeated electronic scattering between nested
regions of the Fermi surface.
Provided the characteristic fluctuation frequency remains much smaller than
the electronic bandwidth, the spin fluctuations may likewise be treated as
an effectively classical random field.

\subsection{Excitonic Insulator}

The present formalism is equally applicable to systems approaching
excitonic condensation \cite{ExI}.
Here the fluctuating order parameter corresponds to a transient electron-hole
pairing amplitude $\Delta\sim\langle c^\dagger_v c_c\rangle$,  
whose temporal fluctuations produce a dynamic pseudogap in precisely the
same manner as the CDW problem. Here the indices $c$ and $v$ denote conduction
and valence band in a semimetal, with band extrema at points of Brillouin zone
linked by nesting vector ${\bf Q}$.

\subsection{Candidate Materials}

Several classes of materials appear particularly well suited for the
application of the present theory.

These include
\begin{itemize}
\item layered transition-metal dichalcogenides close to CDW ordering 
\cite{KNS_12},
\item quasi-one-dimensional Peierls systems \cite{Gruner},
\item chromium and related itinerant antiferromagnets \cite{Mor_85},
\item underdoped cuprate superconductors exhibiting strong charge- or
spin-density-wave fluctuations \cite{Timusk,ufn1,Tamm,PS1,PS2,SK98,Chub_1,Chub_2},
\item excitonic semimetals close to electronic ordering transitions \cite{ExI}.
\end{itemize}
In all these systems short-range order develops over a broad temperature
range before true long-range ordering is established, producing precisely
the fluctuating pseudogap regime described by the present model.

\subsection{Angle-Resolved Photoemission}

The most direct experimental signature of the coherent dynamic regime is
expected in angle-resolved photoemission spectroscopy (ARPES) \cite{ZX}.
According to the present theory the spectral function consists of an
infinite family of dynamically broadened sidebands separated by
$\Delta\omega\sim\omega_0$.
When $\omega_0\gg\Gamma_{\rm eff}$, individual sidebands remain partially 
resolved and should appear as satellite structures accompanying the principal 
quasiparticle peak.

As the effective decoherence increases these satellites broaden and merge,
leading to the familiar pseudogap spectrum characteristic of the
quasistatic fluctuating-gap regime.

\section{Conclusions}

In this work we have formulated a dynamic theory of pseudogap formation
generated by finite-${\bf Q}$ fluctuations possessing finite correlation
time and characteristic oscillation frequency.
Within the commuting (Abelian) approximation to the exact SU(2)
time evolution we derived an explicit double-series representation of the
single-particle Green's function describing coherent multiple scattering
processes. The resulting propagator naturally acquires a generalized
Bogoliubov structure in which every stochastic scattering history is
characterized by an effective dynamic gap with complex Poisson 
distribution, giving an attractive physical picture of dynamic pseudogap 
formation.

A central result is the identification of the
dynamically generated decoherence scale $\Gamma_{\rm eff}$.
The ratio $\omega_0/\Gamma_{\rm eff}$ naturally separates two qualitatively 
different physical regimes. For $\omega_0>\Gamma_{\rm eff}$,
electrons resolve the temporal dynamics of the fluctuating order
parameter and the coherent double-series representation provides a
controlled description of the resulting sideband structure.
Conversely, when $\Gamma_{\rm eff}\gtrsim\omega_0$,
the electron loses phase coherence before resolving the oscillatory
motion of the order parameter.
The dynamic problem then crosses over to an effectively quasistatic
regime described by the exact continued-fraction solution originally
developed for static fluctuating-gap models.
These two descriptions should therefore not be regarded as independent
theories but rather as complementary asymptotic limits of one underlying
dynamic pseudogap problem. The coherent double-series representation and 
the exact continued-fraction recursion derived in the present work should 
therefore be viewed as two complementary nonperturbative descriptions of 
the same dynamic pseudogap problem, with the crossover between them governed 
by the emergent decoherence scale $\Gamma_{\rm eff}$.

Several exact results support the Green's function construction developed in the
present work. For commensurate fluctuations at the Fermi surface
($\xi=0$), where the Hamiltonians commute at different times,
the coherent cumulant expansion becomes exact and reproduces the
complete dynamic solution.
In the opposite overdamped limit the exact continued-fraction recursion
provides the correct quasistatic description.
These two exact asymptotic limits strongly constrain the structure of the
general theory and provide substantial justification for the coherent
approximation over a broad physically relevant parameter range.
Actually the agreement of our numerics with well known exact solutions of
this model probably remains the main justification of our heuristic approach.

The principal approximation of the present work consists in neglecting
the non-Abelian contributions generated by the SU(2) structure of the
time-ordered evolution operator.
These corrections are naturally organized by the Magnus expansion and are
expected to become quantitatively important primarily in the crossover
region $\omega_0\sim\Gamma_{\rm eff}$.
At the same time, both analytical arguments and extensive numerical
calculations indicate that the coherent approximation remains remarkably
accurate over a wide range of model parameters, suggesting that higher
Magnus terms mainly provide quantitative rather than qualitative
corrections.

Further analysis of these difficulties is possible, in principle, by the use of
special diagram technique for (pseudo)spin operators proposed in
Refs. \cite{VLP,IK,IS,BKY,BL}, which may become very useful for future developments
of our theory.

Although formulated for classical Gaussian fluctuations, the present
approach provides a natural starting point for more microscopic
descriptions, based on different mechanisms of electron interactions with
specific types of fluctuations at the edge of soft-mode phase transitions. 
The coherent double-series representation derived here may 
therefore be viewed as the simplest member of a broader class of non-Abelian
fluctuating-gap theories describing dynamic electronic ordering phenomena.
The future perspective requires additional and more deep analysis of the role
of higher order contributions to Magnus expansion, which were dropped here
in commuting approximation. This is actually the main limitation of the present 
theory which does not allow the complete description of the crossover region 
from dynamic problem to static fluctuating gap model. At the same time we do
not expect any drastic deviations from the physical picture described here.

The present formulation has been developed for fluctuations with a fixed
nesting vector ${\bf Q}$ and therefore does not explicitly include
spatial fluctuations of the ordering wave vector.
A complete theory combining both temporal and spatial fluctuations
remains an important problem for future investigations.
Nevertheless, the close correspondence between the present dynamic theory
and the earlier static fluctuating-gap models 
\cite{Sad1,Sad2,Won1,Sad3,SadTim,PS1,PS2,SK98,PRB05,PRB07,ufn2,Chub_1,Chub_2}   
suggests a simple phenomenological interpolation.
In the static theories the inverse correlation length
$\xi_{\rm corr}^{-1}$ enters the electronic Green's function through the
effective damping
$v_F/\xi_{\rm corr}$,
where $v_F$ is the Fermi velocity.
This observation naturally motivates the introduction of the combined
decoherence scale
\begin{equation}
\gamma_{\rm eff}
=\gamma+\frac{v_F}{\xi_{\rm corr}},
\label{gam_eff}
\end{equation}
which simultaneously accounts for temporal and spatial loss of
electronic coherence.
Although phenomenological, this relation provides a plausible bridge
between the present theory and future microscopic descriptions including
both types of fluctuations.

Finally, the present work establishes a unified physical picture of
dynamic pseudogap formation continuously connecting coherent dynamic
sideband physics, quasistatic fluctuating-gap behavior, and the
motional-narrowing limit within a single microscopic framework.
We hope that this approach will provide a useful basis for future
extensions incorporating spatial fluctuations, non-Abelian corrections,
and microscopic models of soft collective modes, as well as for
quantitative comparison with angle-resolved photoemission,
tunneling spectroscopy, optical conductivity, and other experimental
probes of fluctuating electronic order.

\subsection{Funding}

This work was supported by ongoing institutional funding. No additional grants
to carry out this particular research were obtained.

\subsection{Conflict of interests}

The authors of this work declare that they have no conflict of interest.

\subsection{Acknowledgements}

The authors are grateful to an anonymous referee for constructive criticism and
valuable suggestions to imrove the initial version of the paper.


\appendix


\section{Keldysh formalism}
\label{app:Keld}

The time dependence of the fluctuating order parameter makes the conventional
equilibrium Green's function formalism rather limited. Instead, the natural
framework is provided by the non-equilibrium Keldysh technique, in which the
electron propagator is defined on the closed real-time contour. Below we show
how our dynamic pseudogap problem can be reformulated using Keldysh formalism.

We shall also present the general derivation of the dynamic double-series 
Green's function used in the main part of the text directly from the Keldysh 
contour formulation.
The purpose of this derivation is to make explicit how the stochastic history
indices $(n,m)$ and their complex Poisson weights arise from the underlying
Gaussian scattering diagrams, and how the finite-$Q$ Nambu structure produces
the generalized Bogoliubov denominator.

\subsection{Microscopic Hamiltonian and Nambu representation}

The contour-ordered Green's function is defined by
\begin{equation}
G(1,2)
=
-i
\left<
T_C
\psi(1)
\psi^\dagger(2)
\right>,
\label{ContourGreen}
\end{equation}
where $T_C$ denotes ordering along the Keldysh contour.

Introducing Nambu spinors
\begin{equation}
\hat\Psi_{\bf k}(t)
=
\begin{pmatrix}
c_{\bf k}(t)
\\
c_{\bf k+Q}(t)
\end{pmatrix},
\end{equation}
the Hamiltonian assumes the compact matrix form
\begin{equation}
H(t)
=
\sum_{\bf k}
\hat\Psi_{\bf k}^\dagger
\hat H(t)
\hat\Psi_{\bf k},
\end{equation}
with
\begin{equation}
\hat H(t)
=
\xi_k\sigma_3
+
\Delta(t)\sigma_+
+
\Delta^*(t)\sigma_-,
\label{MatrixHamiltonian}
\end{equation}
where $\sigma_i$ are the Pauli matrices.
The important point is that $\sigma_i$ act only in the two-dimensional
$\mathbf k,\mathbf{k+Q}$ space. They are ordinary c-number matrices and the
underlying particles remain canonical fermions.

The Keldysh generating functional is
\begin{equation}
Z
=
\int
{\cal D}
(\hat\Psi^{\dagger},\hat\Psi)
{\cal D}\Delta
\,
{\cal P}[\Delta]
\,
e^{iS[\hat\Psi^{\dagger},\hat\Psi,\Delta]},
\label{GeneratingFunctional}
\end{equation}
where
\begin{equation}
S
=
S_0
+
S_{\rm int}
\end{equation}
is the action on the Keldysh contour.

The electronic part is
\begin{equation}
S_0
=
\int_C
dt
\sum_{\bf k}
\hat\Psi_{\bf k}^{\dagger}
\left(
i\partial_t
-
\xi_{\bf k}\sigma_3
\right)
\hat\Psi_{\bf k},
\end{equation}
while
\begin{align}
S_{\rm int}
=
-
\int_C
dt
\sum_{\bf k}
\hat\Psi_{\bf k}^{\dagger}
\left[
\Delta(t)\sigma_+
+
\Delta^*(t)\sigma_-
\right]
\hat\Psi_{\bf k}.
\label{InteractionAction}
\end{align}
Since the fluctuating field is Gaussian, the functional integration over
$\Delta(t)$ can be carried out exactly. Using the standard Gaussian identity
\begin{align}
&
\int
{\cal D}\Delta
\,
\exp
\left[
-
\Delta^\dagger
K
\Delta
+
J^\dagger\Delta
+
\Delta^\dagger J
\right]
\nonumber\\
&=
(\det K)^{-1}
\exp
\left(
J^\dagger
D
J
\right),
\label{GaussianIdentity}
\end{align}
where $D=K^{-1}$ is time correlation function along the Keldysh contour,
one immediately obtains
\begin{equation}
Z
=
\int
{\cal D}
(\hat\Psi^{\dagger},\hat\Psi)
\,
e^{iS_{\rm eff}},
\label{EffectiveFunctional}
\end{equation}
with the exact effective electronic action
\begin{equation}
S_{\rm eff}
=
S_0
+
S_{\rm ind},
\end{equation}
where
\begin{align}
S_{\rm ind}
=
&
-
\int_C
dt
dt'
\,
D(t-t')
\nonumber\\
&
\times
\sum_{\bf kk'}
\left[
\hat\Psi_{\bf k}^{\dagger}(t)\sigma_+\hat\Psi_{\bf k}(t)
\right]
\left[
\hat\Psi_{\bf k'}^{\dagger}(t')\sigma_-\hat\Psi_{\bf k'}(t')
\right].
\label{EffectiveAction}
\end{align}
Now the fluctuating order parameter has disappeared completely, being replaced
by a retarded nonlocal electron -- electron interaction mediated by the
correlation function of the stochastic field.

Several important observations follow immediately.
First, the effective interaction is nonlocal in time.
Unlike the static theory, the electron retains memory of its previous
scattering events over a time interval determined by the correlation function
$D(t)$.
Second, all information concerning the fluctuating order parameter enters only
through its two-time correlation function.
This remarkable simplification follows directly from the Gaussian nature of the
stochastic process. Finally, Eq.~(\ref{EffectiveAction}) is exact.
No perturbative approximation has yet been introduced.
The only assumption employed thus far is the Gaussian statistics of the
fluctuating order parameter.

Throughout the present work the fluctuating pseudogap field has been
treated as a classical Gaussian stochastic process characterized by the
correlation function
\begin{equation}
\left<
\Delta(\mathbf r,t)
\Delta^*(\mathbf r',t')
\right>
=
D(\mathbf r-\mathbf r',t-t').
\label{K1}
\end{equation}
In the simplified model considered in the main text the spatial correlations
were neglected.

We have seen above that assuming fluctuating field Gaussian, we may integrate
it out exactly to obtain effective action (\ref{EffectiveAction}), where
\begin{equation}
D_C(t_1,t_2)
=
-i
\left<
T_C
\Delta(t_1)
\Delta^*(t_2)
\right>
\label{K6}
\end{equation}
is the contour propagator of the fluctuating order parameter.
The contour Green's function may be decomposed into the standard Keldysh
components,
\begin{equation}
D_C
\Longrightarrow
\left(
D^R,
D^A,
D^K
\right),
\end{equation}
where $D^R$ and $D^A$ describe the causal response of the order
parameter, while $D^K$ determines the fluctuation spectrum.

The present model corresponds to the classical limit in which the order
parameter is treated as an external stochastic field.
In equilibrium we have fluctuation-dissipation theorem:
\begin{equation}
D^K(\omega)=[D^R(\omega)-D^A(\omega)]\coth\left(\frac{\omega}{2T}\right)
\label{FDT}
\end{equation}
so that for $T\gg\omega$ we have $\coth(\omega/2T)\approx 2T/\omega\gg 1$
and $D^K(\omega)\gg D^{R,A}(\omega)$.
In this case the Keldysh component
\begin{equation}
D^K(t)
\propto
\left<
\Delta(t)\Delta^*(0)
\right>,
\label{K7}
\end{equation}
dominates and governs electron dynamics, whereas the retarded and advanced
components actually do not play any role due to the classical nature of our
random field. Actually the exponential damping of the electronic Green's function
does not arise from phenomenological broadening but from averaging over
all realizations of the fluctuating pseudogap field. 

\subsection{Gaussian field contractions}

Introduce the Keldysh contour $C$ and define
\begin{equation}
\hat G_C(1,2)
=
-i
\left\langle
T_C
\hat\Psi(1)\hat\Psi^\dagger(2)
\right\rangle_{\Delta},
\end{equation}
where the average is over the classical fluctuating field.

For a fixed realization of $\Delta(t)$,
\begin{equation}
\left[
i\partial_{t_1}
-
\hat H(t_1)
\right]
\hat G_C(1,2)
=
\delta_C(1-2).
\label{DysonEquationContour}
\end{equation}
Define
\begin{equation}
\hat G_{0C}^{-1}
=
i\partial_t-\xi\sigma_3.
\end{equation}
The contour Dyson equation is therefore
\begin{equation}
\hat G_C
=
\hat G_{0C}
+
\hat G_{0C}\hat V\hat G_C,
\label{ContourDyson}
\end{equation}
with
\begin{equation}
\hat V(t)
=
\Delta(t)\sigma_+
+
\Delta^*(t)\sigma_-.
\end{equation}
Iterating Eq.~(\ref{ContourDyson}) gives
\begin{widetext}
\begin{align}
\hat G_C
={}&
\hat G_{0C}
+
\hat G_{0C}\hat V\hat G_{0C}
+
\hat G_{0C}\hat V\hat G_{0C}\hat V\hat G_{0C}
\nonumber\
+
\hat G_{0C}\hat V\hat G_{0C}\hat V
\hat G_{0C}\hat V\hat G_{0C}
+\cdots .
\label{DysonSeries}
\end{align}
\end{widetext}
This is the starting point for the Keldysh diagrammatic expansion.

The classical field has zero mean,
\begin{equation}
\langle\Delta(t)\rangle=0,
\end{equation}
and correlation function
\begin{equation}
D_C(t,t')
=
\left\langle
\Delta(t)\Delta^*(t')
\right\rangle_C.
\end{equation}
For the incommensurate case,
\begin{equation}
\left\langle
\Delta(t)\Delta(t')
\right\rangle
=
0,
\end{equation}
while
\begin{equation}
D(t-t')
=
\Delta^2
e^{-\gamma|t-t'|}
\cos[\omega_0(t-t')].
\label{Dcorrelation}
\end{equation}
The Gaussian nature of the classical field means that its higher moments are
reduced by ordinary Gaussian Wick contraction,
\begin{equation}
\langle
\Delta_1\Delta_2
\Delta_3^*\Delta_4^*
\rangle
=
D_{12}D_{34}
+
D_{14}D_{32},
\end{equation}
and similarly at higher order.
It is simply Wick's theorem for the classical Gaussian random field.
Every pair of field vertices in the electronic Dyson series is therefore joined
by a contour disorder line $D_C$.

The retarded electronic Green's function is given by
\begin{equation}
\hat G^R(\epsilon,\xi)
=\left[
\epsilon+i0-\xi\sigma_3
-\hat\Sigma^R(\epsilon,\xi)
\right]^{-1}.
\label{K_Dyson}
\end{equation}
The self-energy generated by the classical field has the schematic
Keldysh form
\begin{eqnarray}
\hat\Sigma^R
=
\int\frac{d\Omega}{2\pi}
\left[
D^K(\Omega)
\sigma_a
\hat G^R(\epsilon-\Omega,\xi)
\sigma_a\right.
+\nonumber\\
+\left.D^R(\Omega)
\sigma_a
\hat G^K(\epsilon-\Omega,\xi)
\sigma_a
\right].
\label{K_selfenergy}
\end{eqnarray}
For a prescribed classical stochastic field the dominant noise
contribution may be represented as
\begin{equation}
\hat\Sigma^R_{\rm cl}
\simeq
\int\frac{d\Omega}{2\pi}
D^K(\Omega)
\sigma_a
\hat G^R(\epsilon-\Omega,\xi)
\sigma_a,
\label{classical_selfenergy}
\end{equation}
Equation~(\ref{classical_selfenergy}) displays the fundamental
separation of variables.  The noise spectrum is independent of $\xi$,
whereas the electronic propagator appearing inside the convolution
depends explicitly on $\xi$.

\subsection{The finite-Q pseudospin structure of the diagrams}

The free propagator is
\begin{equation}
\hat G_{0C}^{-1}(\omega)
=
\omega\sigma_0-\xi\sigma_3.
\end{equation}
Consequently,
\begin{equation}
\hat G_{0C}(\omega,\xi)
=
\frac{
\omega\sigma_0+\xi\sigma_3
}{
\omega^2-\xi^2
}.
\end{equation}
The scattering vertices are off diagonal:
\begin{equation}
V_+=\sigma_+,
\qquad
V_-=\sigma_-.
\end{equation}
Using
\begin{equation}
\sigma_\pm\sigma_3
=
\mp\sigma_3\sigma_\pm,
\end{equation}
one obtains
\begin{equation}
\sigma_\pm
\hat G_0(\xi)
=
\hat G_0(-\xi)\sigma_\pm.
\label{BranchFlip}
\end{equation}
Thus every scattering event changes the branch,
\begin{equation}
\xi
\longrightarrow
-\xi
\longrightarrow
\xi
\longrightarrow
-\xi
\cdots .
\end{equation}
This branch alternation is the microscopic origin of the Nambu -- Bogoliubov
structure. It is already present before any stochastic averaging is performed.

For positive time difference,
\begin{equation}
D(t)
=
\frac{\Delta^2}{2}
e^{-\gamma t}
\left(
e^{-i\omega_0t}
+
e^{+i\omega_0t}
\right).
\end{equation}
It is convenient to introduce two elementary dynamic propagators,
\begin{equation}
D_+(t)
=
\frac{\Delta^2}{2}
e^{-(\gamma+i\omega_0)t},
\end{equation}
and
\begin{equation}
D_-(t)
=
\frac{\Delta^2}{2}
e^{-(\gamma-i\omega_0)t}.
\end{equation}
The complete diagram series can consequently be classified according to the
number of $D_+$ and $D_-$ lines occurring in a scattering history.
We denote these numbers by
\begin{equation}
n=0,1,2,\ldots ,
\qquad
m=0,1,2,\ldots .
\end{equation}
The pair $(n,m)$ is therefore a dynamic history index.

Consider first a sequence of $n$ identical $D_+$ insertions. Because each
$D_+$ is exponential, the ordered contour integral has the generic form
\begin{equation}
I_n^{(+)}(t)
=
\int_{0<t_n<\cdots<t_1<t}
dt_1\cdots dt_n,
\prod_{j=1}^{n}
e^{-\Omega_+ (t_j-t_{j+1})},
\end{equation}
where
\begin{equation}
t_{n+1}=0,
\qquad
\Omega_+=\gamma+i\omega_0.
\end{equation}
Introduce the interval variables
\begin{equation}
\tau_j=t_j-t_{j+1}.
\end{equation}
The ordered domain becomes
\begin{equation}
\tau_j\ge0,
\qquad
\sum_{j=1}^{n}\tau_j\le t.
\end{equation}
The integral is consequently generated recursively. Defining
\begin{equation}
A_n^{(+)}
\end{equation}
as the coefficient of the corresponding history, one obtains
\begin{equation}
A_n^{(+)}
=
\frac{\Delta^2/2}{\Omega_+}
\frac{1}{n}
A_{n-1}^{(+)}.
\label{OneHistoryRecurrence}
\end{equation}
Hence
\begin{equation}
A_n^{(+)}
=
\frac{1}{n!}
\left(
\frac{\Delta^2}
{2\Omega_+^2}
\right)^n.
\end{equation}
The same construction for the opposite oscillatory component gives
\begin{equation}
A_m^{(-)}
=
\frac{1}{m!}
\left(
\frac{\Delta^2}
{2\Omega_-^2}
\right)^m,
\end{equation}
where
\begin{equation}
\Omega_-=\gamma-i\omega_0.
\end{equation}
Thus the factorials in the Poisson weights arise directly from the ordered
Keldysh time integrations.

For a general $(n,m)$ sector, adding one additional $D_+$ or $D_-$ line gives
\begin{equation}
A_{nm}
=
\frac{\Delta^2}{2}
\frac{
A_{n-1,m}/\Omega_+
+
A_{n,m-1}/\Omega_-
}{
n\Omega_++m\Omega_-}.
\label{HistoryRecurrence}
\end{equation}
The boundary condition is
\begin{equation}
A_{00}=1,
\end{equation}
and
\begin{equation}
A_{n<0,m}=A_{n,m<0}=0.
\end{equation}
The solution of Eq.~(\ref{HistoryRecurrence}) is
\begin{equation}
A_{nm}
=
e^{-\lambda_1-\lambda_2}
\frac{\lambda_1^n}{n!}
\frac{\lambda_2^m}{m!},
\label{PoissonHistory}
\end{equation}
where $\lambda_1$ and $\lambda_2$ were defined in Eq. (\ref{lambi})
of the main text.
The exponential prefactor follows from the sum of disconnected repetitions
of the elementary histories.

Thus the entire Gaussian Keldysh diagram series has been reorganized into
\begin{equation}
\sum_{n,m}
P_{\lambda_1}(n)
P_{\lambda_2}(m)
[\text{Nambu propagator of history }(nm)]
\end{equation}
The procedure described here presents an alternative derivation of
our Poisson averaged histories approach used in the main part of the text.

This derivation implicitly assumed that diagrams differing only by
permutations of equivalent $D_+$ and $D_-$ insertions can be grouped into
the same history sector.
In the exact finite-$Q$ theory this is not generally allowed.
For example, consider two successive transverse scattering operations.
Their two possible orderings contain
\begin{equation}
\sigma_+\sigma_-
\end{equation}
and
\begin{equation}
\sigma_-\sigma_+,
\end{equation}
whose difference is
\begin{equation}
[\sigma_+,\sigma_-]=\sigma_3.
\end{equation}
The Abelian approximation amounts to dropping precisely this ordering-sensitive
difference.
Diagrammatically,
\begin{equation}
D_+D_-
\quad\leftrightarrow\quad
D_-D_+
\end{equation}
are assigned the same history amplitude.

The approximation therefore concerns the
\emph{ordering of the pseudospin vertices}, not the stochastic Gaussian
contractions themselves.

\subsection{History Hamiltonian}

After the Gaussian diagrams have been grouped into an $(n,m)$ sector, the
remaining electronic object is a two-component propagator.
Consider the commensurate case. The dynamic phase accumulated by the history is
\begin{equation}
\Omega_{nm}
=
(n-m)\omega_0-i(n+m)\gamma.
\end{equation}
The nonperturbative damping generated by the repeated dynamic scattering is
denoted by
\begin{equation}
\Gamma_{\mathrm{eff}}
=
\frac{\gamma\Delta^2}
{\omega_0^2+\gamma^2}.
\end{equation}
Within the Abelian history construction the corresponding complex transverse
gap is
\begin{equation}
\Delta_{nm}
=
(n-m)\omega_0
-i(n+m)\gamma
-i\Gamma_{\mathrm{eff}}.
\label{HistoryGap}
\end{equation}
Equivalently,
\begin{equation}
\Delta_{nm}
=
(\omega_0-i\gamma)(n-\lambda_1)
+
(-\omega_0-i\gamma)(m-\lambda_2),
\end{equation}
The history Hamiltonian is therefore assumed to be
\begin{equation}
\hat H_{nm}
=
\xi\sigma_3+
\Delta_{nm}\sigma_1.
\label{HistoryHamilt}
\end{equation}
This step is the finite-$Q$ part of the construction: the dynamic history
does not produce a scalar shift of $\xi$. It produces a transverse matrix
element connecting the two nested electronic branches.

The Green's function associated with a fixed history for commensurate 
case satisfies
\begin{equation}
\left[
\omega\sigma_0-\hat H_{nm}
\right]
\hat G_{nm}
=
\sigma_0.
\end{equation}
Using Eq.~(\ref{HistoryHamilt}),
\begin{equation}
\left[
\omega\sigma_0
-\xi\sigma_3
-\Delta_{nm}\sigma_1
\right]
\hat G_{nm}
=
\sigma_0.
\end{equation}
Since
\begin{equation}
\{\sigma_1,\sigma_3\}=0,
\end{equation}
we have
\begin{equation}
(\xi\sigma_3+\Delta_{nm}\sigma_1)^2
=
\xi^2+\Delta_{nm}^2.
\end{equation}
Consequently,
\begin{equation}
\hat G_{nm}
=
\frac{
\omega\sigma_0+
\xi\sigma_3+
\Delta_{nm}\sigma_1
}{
\omega^2-\xi^2-\Delta_{nm}^2
}.
\label{HistoryGreen}
\end{equation}
The physical electron component is
\begin{equation}
G_{nm,11}
(\omega,\xi)
=
\frac{
\omega+\xi
}{
\omega^2-\xi^2-\Delta_{nm}^2
}.
\label{HistoryElectronGreen}
\end{equation}
Thus the pseudogap appears directly from the off-diagonal finite-$Q$
scattering vertex.

Combining the Gaussian history amplitude Eq.~(\ref{PoissonHistory}) with the
history-resolved Nambu propagator Eq.~(\ref{HistoryGreen}), we obtain
\begin{equation}
\hat G
(\omega,\xi)
=
e^{-\lambda_1-\lambda_2}
\sum_{n,m=0}^{\infty}
\frac{\lambda_1^n}{n!}
\frac{\lambda_2^m}{m!}
\frac{
\omega\sigma_0+
\xi\sigma_3+
\Delta_{nm}\sigma_1
}{
\omega^2-\xi^2-\Delta_{nm}^2
}.
\label{FullAbelianMatrixGF}
\end{equation}
The physical electron Green function is therefore
\begin{equation}
G
(\omega,\xi)
=
e^{-\lambda_1-\lambda_2}
\sum_{n,m=0}^{\infty}
\frac{\lambda_1^n}{n!}
\frac{\lambda_2^m}{m!}
\frac{
\omega+\xi
}{
\omega^2-\xi^2-\Delta_{nm}^2
}.
\label{FullAbelianGF}
\end{equation}
This expression is obtained here directly by reorganizing the Keldysh Dyson
series and its Gaussian field contractions.

The structure of Eq.~(\ref{FullAbelianGF}) can be understood directly from
the diagrams.
At zero scattering order,
\begin{equation}
G^{(0)}
=
\frac{1}{\omega-\xi+i0}.
\end{equation}
At second order, the elementary contribution contains
\begin{equation}
G_0(\xi)
,
\sigma_+
G_0(-\xi)
,
\sigma_-
G_0(\xi)
\end{equation}
multiplied by one field contraction $D_C$.
At fourth order one obtains
\begin{align}
G^{(4)}
\sim{}&
G_0
\sigma_+
G_0
\sigma_-
G_0
\sigma_+
G_0
\sigma_-
G_0
\left[
D_{12}D_{34}
+
D_{14}D_{32}
\right].
\end{align}
The two Gaussian pairings correspond to the two topologically distinct
contractions.

At arbitrary order, the Gaussian contractions generate all pairings of the
field vertices. The Abelian approximation identifies all diagrams having
the same numbers $(n,m)$ of the two elementary dynamical components.
Their sum is precisely represented by the history amplitude
\begin{equation}
P_{\lambda_1}(n)P_{\lambda_2}(m).
\end{equation}
Thus Eq.~(\ref{FullAbelianGF}) is not a phenomenological ansatz: it is the
compact representation of the Abelianized infinite Keldysh diagram series.

For the commensurate case,
\begin{equation}
\Delta=\Delta^*,
\end{equation}
and the construction above directly gives the Abelian history Hamiltonian
\begin{equation}
\hat H_{nm}
=
\xi\sigma_3+
\Delta_{nm}\sigma_1.
\end{equation}
At $\xi=0$,
\begin{equation}
\hat H(t)=\Delta(t)\sigma_1,
\end{equation}
so
\begin{equation}
[\hat H(t_1),\hat H(t_2)]=0
\end{equation}
identically. The Abelian construction is therefore exact in this limit.

For the incommensurate case,
\begin{equation}
\Delta(t)=\Delta_1(t)+i\Delta_2(t),
\end{equation}
and
\begin{equation}
\hat H(t)
=
\xi\sigma_3+
\Delta_1(t)\sigma_1+
\Delta_2(t)\sigma_2.
\end{equation}
The corresponding Abelian history construction introduces
\begin{equation}
\hat H_{nm}
=
\xi\sigma_3+
\Delta_{1,n}\sigma_1+
\Delta_{2,m}\sigma_2,
\end{equation}
with
\begin{equation}
\Delta_{1,n}
=
(\omega_0-i\gamma)(n-\lambda_1),
\end{equation}
\begin{equation}
\Delta_{2,m}
=
(-\omega_0-i\gamma)(m-\lambda_2).
\end{equation}
Hence
\begin{equation}
\hat H_{nm}^2
=
\left[
\xi^2+
\Delta_{1,n}^2+
\Delta_{2,m}^2
\right]\sigma_0,
\end{equation}
and
\begin{equation}
G_{nm,11}
=
\frac{\omega+\xi}
{
\omega^2-\xi^2
-\Delta_{1,n}^2
-\Delta_{2,m}^2
}.
\end{equation}
The Gaussian history weights remain
\begin{equation}
P_{\lambda_1}(n)P_{\lambda_2}(m).
\end{equation}
For this incommensurate construction, however, one must explicitly state that
the matrix history assignment is an Abelian $O(2)$ extension rather than an
exact solution of the full non-Abelian Keldysh problem.

The discarded contributions can be identified explicitly.
Consider two transverse vertices,
\begin{equation}
V_+=\Delta\sigma_+,
\qquad
V_-=\Delta^*\sigma_-.
\end{equation}
The two possible orderings differ by
\begin{equation}
V_+V_--V_-V_+
=
|\Delta|^2
[\sigma_+,\sigma_-]
=
|\Delta|^2\sigma_3.
\end{equation}
In the exact Keldysh series these two orderings correspond to distinct
pseudospin histories.
The Abelian construction replaces
\begin{equation}
V_+V_-
\quad\text{and}\quad
V_-V_+
\end{equation}
by their common commuting history contribution.
Thus the approximation is precisely
\begin{equation}
[\sigma_+,\sigma_-]\rightarrow0
\end{equation}
inside the history resummation.
Importantly, we do \emph{not} set
\begin{equation}
\sigma_\pm\rightarrow0.
\end{equation}
The transverse scattering vertices remain and are responsible for the pseudogap.
This observation provides the natural starting point for a systematic
non-Abelian extension.


\section{Commensurate and Incommensurate Fluctuations}
\label{app:Incomm_vs_Comm}

\subsection{Real fluctuating order parameter}

For a commensurate charge-density wave the order parameter may be chosen real,
and the interaction Hamiltonian takes the form
\begin{equation}
H_{\rm int}(t)
=
\Delta(t)\sigma_1,
\label{HintReal}
\end{equation}
The perturbation expansion therefore involves only a single fluctuating
Gaussian field with correlation function
\begin{equation}
\left<
\Delta(t_1)\Delta(t_2)
\right>
=
D(t_1-t_2).
\end{equation}
All interaction vertices are identical, and every Wick contraction involves
the same scalar propagator.

The square of interaction Hamiltonian reduces to:
\begin{equation}
H_{\rm int}^2
=
\Delta^2\,\mathbf{1},
\end{equation}
leading to the single effective history-dependent gap discussed in the main
text.

\subsection{Complex fluctuating order parameter}

For an incommensurate density wave the order parameter is intrinsically
complex,
\begin{equation}
\Delta
=
\Delta_1+i\Delta_2,
\end{equation}
and the interaction Hamiltonian may be written as
\begin{equation}
H_{\rm int}(t)
=
\Delta_1(t)\sigma_1
+
\Delta_2(t)\sigma_2.
\label{HintComplex}
\end{equation}
The perturbation expansion now contains two statistically independent
Gaussian fields.

Assuming isotropic Gaussian fluctuations, the elementary correlator becomes
\begin{equation}
\left<
\Delta_a(t_1)
\Delta_b(t_2)
\right>
=
\delta_{ab}
D(t_1-t_2),
\qquad
a,b=1,2.
\label{TensorCorrelator}
\end{equation}
Equation~(\ref{TensorCorrelator}) immediately implies that Wick contractions
preserve the component index of the fluctuating field.
Mixed contractions between $\Delta_1$ and $\Delta_2$ vanish identically.

Consequently, the perturbation expansion naturally separates into two
independent contraction families.
Within the dynamic double-series representation these two families are
naturally labelled by the stochastic history indices $n$ and $m$.

The decomposition employed above follows directly from the standard
electron--density-wave interaction Hamiltonian. For an incommensurate charge-
density wave the interaction with the fluctuating order parameter is naturally
written as
\begin{equation}
H_{\rm int}(t)
=
\Delta(t)\sigma_{+}
+
\Delta^{*}(t)\sigma_{-},
\label{HintSigmaPM}
\end{equation}
Writing the complex order parameter as
\begin{equation}
\Delta
=
\Delta_1+i\Delta_2,
\qquad
\Delta^{*}
=
\Delta_1-i\Delta_2,
\end{equation}
one immediately obtains
\begin{eqnarray}
H_{\rm int}
&=&
(\Delta_1+i\Delta_2)\sigma_{+}
+
(\Delta_1-i\Delta_2)\sigma_{-}
\nonumber\\
&=&
\Delta_1(\sigma_{+}+\sigma_{-})
+
i\Delta_2(\sigma_{+}-\sigma_{-}).
\end{eqnarray}
Using the identities
\begin{equation}
\sigma_{+}+\sigma_{-}
=
\sigma_1,
\qquad
i(\sigma_{+}-\sigma_{-})
=
-\sigma_2,
\end{equation}
one finally arrives at
\begin{equation}
H_{\rm int}
=
\Delta_1\sigma_1
-
\Delta_2\sigma_2.
\label{HintXY}
\end{equation}
The relative sign depends only upon the convention chosen for the definition
of $\sigma_{\pm}$ or, equivalently, for the phase of the complex order
parameter, and therefore has no physical significance. The important point is
that the two real components of the complex order parameter couple to two
mutually anticommuting Pauli matrices.

The square of the interaction Hamiltonian therefore becomes
\begin{eqnarray}
H_{\rm int}^2
&=&
\Delta_1^2\sigma_1^2
+
\Delta_2^2\sigma_2^2
-
\Delta_1\Delta_2
(\sigma_1\sigma_2+\sigma_2\sigma_1).
\end{eqnarray}
Using
\begin{equation}
\sigma_1^2
=
\sigma_2^2
=
\mathbf 1,
\qquad
\{\sigma_1,\sigma_2\}=0,
\end{equation}
one obtains
\begin{equation}
H_{\rm int}^2
=
(\Delta_1^2+\Delta_2^2)\mathbf1.
\label{HintSquare}
\end{equation}
Equation~(\ref{HintSquare}) demonstrates that the spectrum of our model
depends only upon the rotationally invariant quadratic form
$\Delta^2=\Delta_1^2+\Delta_2^2$. 

\subsection{Possible generalizations}

The above construction suggests that the dynamic double-series formalism is
largely independent of the internal symmetry of the fluctuating order
parameter. Instead, the symmetry enters through the matrix algebra associated
with the interaction Hamiltonian.

For example, a three-component fluctuating order parameter may formally be
written as
\begin{equation}
H_{\rm int}
=
\sum_{a=1}^{3}
\Delta_a\sigma_a,
\end{equation}
which immediately gives
\begin{equation}
H_{\rm int}^2
=
\sum_{a=1}^{3}
\Delta_a^2\,\mathbf1.
\end{equation}
This observation suggests that the history Green's function may generally be
constructed from the invariant quadratic form associated with the internal
symmetry of the fluctuating order parameter. In this sense, the dynamic
double-series representation appears to be universal, while different physical
systems correspond to different realizations of the underlying algebra.

The above considerations are intended only as an algebraic interpretation of
the proposed history Green's functions. A complete microscopic derivation for
general multicomponent order parameters remains an interesting subject for
future investigation.

\section{Magnus Expansion and non -- Abelian Stochastic SU(2) Evolution}
\label{sec:SU2}

\subsection{Time evolution and Magnus terms}

The exact time evolution operator is
\begin{equation}
U(t)
=
T
\exp
\left[
-i
\int_0^t
dt'
\,H(t')
\right].
\end{equation}
Interaction operators at different times in general do not
commute,
\begin{equation}
[H(t_1),H(t_2)]
\neq0.
\end{equation}

The evolution operator may now be written as
\begin{equation}
U(t)
=
\exp
\left[
\Omega_1
+
\Omega_2
+
\Omega_3
+\cdots
\right],
\end{equation}
where
\begin{equation}
\Omega_1
=
-i
\int_0^t
dt_1
\,H(t_1)
\end{equation}
is the Abelian contribution.

The leading non-Abelian correction is
\begin{equation}
\Omega_2
=
-\frac12
\int_0^t
dt_1
\int_0^{t_1}
dt_2
\,
[H(t_1),H(t_2)].
\label{Omega2}
\end{equation}

\subsection{Commensurate versus Incommensurate Order}
\label{app:MagnusIncommensurate}

An important qualitative difference between commensurate and
incommensurate dynamic pseudogap fluctuations appears already at the
level of the Magnus expansion.

For the commensurate model discussed in the main text, the Nambu
Hamiltonian is
\begin{equation}
\hat H(t)
=
\xi\sigma_3
+
\Delta(t)\sigma_1 ,
\label{A1}
\end{equation}
where the fluctuating field $\Delta(t)$ is real.

The second Magnus term is controlled by
\begin{equation}
[\hat H(t_1),\hat H(t_2)]
=
2i\xi
\left[
\Delta(t_2)-\Delta(t_1)
\right]
\sigma_2 .
\label{A2}
\end{equation}
Therefore
\begin{equation}
\xi=0
\qquad\Longrightarrow\qquad
[\hat H(t_1),\hat H(t_2)]=0 ,
\label{A3}
\end{equation}
so that all higher Magnus terms vanish identically,
\begin{equation}
\Omega_2=\Omega_3=\cdots=0.
\end{equation}
Consequently, at the Fermi surface the cumulant solution becomes exact,
because the Hamiltonians at different times commute.

The static contribution $\xi\sigma_3$ acts as a pseudomagnetic field directed 
along the third axis of SU(2) space, whereas $\Delta(t)\sigma_1$ 
represents a time-dependent field directed along the first axis.
Because these two pseudomagnetic fields are not parallel, successive
infinitesimal rotations occur about different axes.
The resulting evolution depends on the order in which the rotations are
performed, producing the Magnus commutators.
At the Fermi surface the static component disappears, leaving only the
time-dependent field along a single SU(2) axis.
All infinitesimal rotations then commute identically.

For incommensurate order the situation changes qualitatively.
The Hamiltonian is now
\begin{equation}
\hat H(t)
=
\xi\sigma_3
+
\Delta(t)\sigma_+
+
\Delta^*(t)\sigma_- ,
\label{A4}
\end{equation}
or equivalently
\begin{equation}
\hat H(t)
=
\xi\sigma_3
+
\Delta_1(t)\sigma_1
+
\Delta_2(t)\sigma_2 ,
\label{A5}
\end{equation}
where
\begin{equation}
\Delta(t)=\Delta_1(t)+i\Delta_2(t).
\end{equation}
Then one obtains
\begin{align}
[\hat H(t_1),\hat H(t_2)]
=
&
\,2i\xi
\left[
\Delta_1(t_2)-\Delta_1(t_1)
\right]
\sigma_2
\nonumber\\
&
-
2i\xi
\left[
\Delta_2(t_2)-\Delta_2(t_1)
\right]
\sigma_1
\nonumber\\
&
+
2i
\Bigl[
\Delta_1(t_1)\Delta_2(t_2)
-
\Delta_2(t_1)\Delta_1(t_2)
\Bigr]
\sigma_3 .
\label{A6}
\end{align}
Unlike the commensurate case, the last contribution survives even
exactly at the Fermi surface ($\xi=0$), giving
\begin{equation}
[\hat H(t_1),\hat H(t_2)]
=
2i
\Bigl[
\Delta_1(t_1)\Delta_2(t_2)
-
\Delta_2(t_1)\Delta_1(t_2)
\Bigr]
\sigma_3 .
\label{A7}
\end{equation}
Therefore
\begin{equation}
\Omega_2\neq0
\qquad
(\xi=0)
\end{equation}
for generic incommensurate fluctuations.

Equation (\ref{A7}) admits a simple geometric interpretation.
Introducing the two-dimensional vector order parameter as:
\begin{equation}
\bm{\Delta}(t)
=
\left(
\Delta_1(t),
\Delta_2(t)
\right),
\end{equation}
the commutator may be written as
\begin{equation}
[\hat H(t_1),\hat H(t_2)]
=
2i
\left[
\bm{\Delta}(t_1)
\times
\bm{\Delta}(t_2)
\right]
\sigma_3 .
\label{A8}
\end{equation}
The second Magnus term is therefore proportional to the oriented area
spanned by the order parameter during its stochastic evolution at two different
times in the complex order-parameter plane.

This result demonstrates that the noncommutativity of the
incommensurate problem has a purely geometric origin.
Although the fluctuating field is classical, the pseudospin Hamiltonian
rotates about different axes in SU(2) space at different times.
Consequently, the time-evolution operators do not commute even at the
Fermi surface.

By contrast, in the commensurate model the fluctuating field is purely
real, so all Hamiltonians become proportional to the same generator
$\sigma_1$ when $\xi=0$.
The dynamics then becomes effectively Abelian and the Magnus expansion
terminates after the first term.

\section{The limit of $\gamma=0$,\ $\omega_0\to 0$ }
\label{app:Static}

For the case of incommensurate fluctuations (complex order parameter)
the double series for the Green's
functions in the limit of $\gamma=0$ and $\omega_0\ll\Delta$ can be written as:
\begin{widetext}
\begin{equation}
G(\omega,\xi)
=\sum_{n,m=0}^{\infty}
P_{\lambda_1}(n)P_{\lambda_2}(m)
\frac{\omega+\xi}
{\omega^2-\xi^2-\omega_0^2(n-\lambda_1)^2
-\omega_0^2(m-\lambda_2)^2}.
\label{DoubleSerInc}
\end{equation}
\end{widetext}
where $\lambda_i=\Delta^2/2\omega_0^2\gg 1$ and the complex Poisson
distributions reduce to the Gaussian form:
\begin{equation}
P_{\lambda_i}(k)=\frac{\omega_0}{\sqrt{\pi}\Delta}
\exp[-\frac{\omega_0^2(k-\Delta^2/2\omega_0^2)^2}{\Delta^2}].
\end{equation}
Then introducing $x=n-\Delta^2/2\omega_0^2$ and $y=m-\Delta^2/2\omega_0^2$ 
we can write Green's function (\ref{DoubleSerInc}) as double integral:
\begin{eqnarray}
G(\omega,\xi)=\frac{\omega_0^2}{\pi\Delta^2}\int_{-\infty}^{+\infty}dx
\int_{-\infty}^{+\infty}dy\exp(-\omega_0^2\frac{x^2+y^2}{\Delta^2})\times
\nonumber\\
\times\frac{\omega+\xi}{\omega^2-\xi^2-\omega_0^2(x^2+y^2)}
\nonumber\\
\end{eqnarray}
After the obvious transformation to polar coordinates $\rho^2=x^2+y^2$ and
$\phi$, $dxdy\to d\phi\rho d\rho$ we obtain:
\begin{eqnarray}
G(\omega,\xi)=\frac{\omega_0^2}{\pi\Delta^2}\int_{0}^{2\pi}d\phi
\int_{0}^{+\infty}d\rho \rho\exp(-\frac{\omega_0^2}{\Delta^2}\rho^2)\times
\nonumber\\
\times\frac{\omega+\xi}{\omega^2-\xi^2-\omega_0^2\rho^2}\nonumber\\
\end{eqnarray}
Then introducing $W=\rho\omega_0$ we get:
\begin{equation}
G(\omega,\xi)=\int_{0}^{+\infty}dW\frac{2W}{\Delta^2}
\exp(-\frac{W^2}{\Delta^2})
\frac{\omega+\xi}{\omega^2-\xi^2-W^2}
\end{equation}
which describes the well known result \cite{Sad1,Sad2}
for the Green's function of an electron
moving in a periodic field with a random phase and Rayleigh distributed
amplitude.

It can be easily shown that for the case of commensurate fluctuation 
in the same limit of $\gamma=0$ and $\omega_0\ll\Delta$
our double series for the Green's function
\begin{eqnarray}
G(\omega,\xi)
=\sum_{n,m=0}^{\infty}
P_{\lambda_1}(n)P_{\lambda_2}(m)\times\nonumber\\
\times\frac{\omega+\xi}
{\omega^2-\xi^2-[\omega_0(n-\lambda_1)
-\omega_0(m-\lambda_2)]^2},
\label{DoubleSerComm}
\end{eqnarray}
reduces to the simple Gaussian average expression \cite{Won1}:
\begin{equation}
G(\omega,\xi)=\int_{-\infty}^{+\infty}dW\frac{1}{\sqrt{2\pi}\Delta}
\exp(-\frac{W^2}{2\Delta^2})
\frac{\omega+\xi}{\omega^2-\xi^2-W^2}
\end{equation}
which corresponds to the case of real order parameter.

\section{Soft Phonon Near CDW Instability and the Origin of
Dynamic Random Field}
\label{app:Microscopic}

In the main text the fluctuating order parameter has been treated as a
classical Gaussian random field with prescribed temporal correlation
function.
The purpose of this Appendix is to discuss the microscopic origin of this
description and to establish its relation to the conventional
electron--phonon Hamiltonian.

Consider the standard electron--phonon interaction,
\begin{equation}
H_{ep}
=
\sum_{{\bf k,q}}
g_{\bf q}
\left(b_{\bf q}+b_{\bf{-q}}^{\dagger}
\right)
c_{\bf k+q}^{\dagger}
c_{\bf k}.
\label{AE1}
\end{equation}
Near a Peierls or charge-density-wave transition the phonon modes with
wave vectors close to nesting vector
${\bf q}\sim {\bf Q}$
soften and at ${\bf q=Q}$ we have
$\omega_{\bf Q}\rightarrow 0$.
Corresponding Bose occupation numbers become large and
\begin{equation}
n_B(\omega_{\bf Q})
=
\frac{1}
{e^{\omega_{\bf Q}/T}-1}
\simeq
\frac{T}{\omega_{\bf Q}},
\end{equation}
diverges as the transition is approached.
The phonon coordinates with ${\bf q\sim Q}$
\begin{equation}
u_{\bf q}\sim u_{\bf Q}
=
\frac{1}{\sqrt{2M\omega_{\bf Q}}}
(b_{\bf Q}+b_{{\bf -Q}}^{\dagger})
\end{equation}
acquire large thermal fluctuations and behave as an effectively
classical stochastic variable.
The interaction Hamiltonian may then be approximately written as
\begin{equation}
H_{ep}
=
\sum_{\bf k}
\Delta(t)
c_{\bf k+Q}^{\dagger}
c_{\bf k}
+
\mathrm{H.c.},
\end{equation}
where
$\Delta(t)
= g_{\bf Q} u_{\bf Q}(t)$.
This is precisely the phenomenological Hamiltonian employed throughout the
present work.

More formally we can consider the usual finite temperature second -- order 
contribution to retarded electron self -- energy due to electron -- phonon
interaction, obtained by the standard  analytical continuation from
Matsubara frequencies  \cite{SadovskiiBook}:
\begin{eqnarray}
\Sigma^R(\omega,{\bf k})=\sum_{\bf q}|g_{\bf q}|^2
\Biggl\{\frac{f_{\bf k+q}+n_B(\omega_{\bf q})}{\omega - \xi_{\bf k+q}
+\omega_{\bf q}+i\delta}\nonumber\\
+ \frac{1-f_{\bf k+q}+n_B(\omega_{\bf q})}
{\omega-\xi_{\bf k+q}-\omega_{\bf q} + i\delta}\Biggr\}
\label{self-energy}
\end{eqnarray}
where $f_{\bf k}$ is Fermi function. 
Now $f_{\bf k}\sim (1-f_{\bf k})\sim 1$, so that in case of $T\gg \omega_{\bf q}$ 
and $n_B(\omega_{\bf q})\sim T/\omega_{\bf q}$ we can rewrite 
(\ref{self-energy}) as:
\begin{eqnarray}
\Sigma^R(\omega,{\bf k})=\sum_{\bf q}|g_{\bf q}|^2\frac{T}{\omega_{\bf q}}
\Biggl\{\frac{1}{\omega - \xi_{\bf k+q}
+\omega_{\bf q}+i\delta}\nonumber\\
+ \frac{1}
{\omega-\xi_{\bf k+q}-\omega_{\bf q} + i\delta}\Biggr\}
\label{self_en}
\end{eqnarray}
For soft -- mode spectrum near Peierls transition we can write:
\begin{equation}
\omega_{\bf q}=\omega_0+c^2(\bf q-Q)^2
\label{softmode}
\end{equation}
so that approximately we have:
\begin{eqnarray}
&&\Sigma^R(\omega,{\bf k})\approx\sum_{\bf q}|g_{\bf q}|^2
\frac{T}{\omega_0+c^2(\bf q-Q)^2}\times
\nonumber\\
&&\times\Biggl\{\frac{1}{\omega+\xi
+\omega_0+i\delta}
+\frac{1}
{\omega+\xi-\omega_0 + i\delta}\Biggr\}
=\nonumber\\
&&=\frac{1}{2}\Delta^2\Biggl\{\frac{1}{\omega+\xi
+\omega_0+i\delta}
+ \frac{1}
{\omega+\xi-\omega_0 + i\delta}\Biggr\},
\label{self_en_random_field}
\end{eqnarray}
where we used nesting condition $\xi_{\bf k}=-\xi_{\bf k-Q}=\xi$ 
and introduced the effective amplitude as:
\begin{equation}
\Delta^2=\sum_{\bf q} |g_{\bf q}|^2\frac{2T}{\omega_0+c^2({\bf q-Q})^2}.
\label{DeltaFluct}
\end{equation}
We see that the soft -- mode at the Peierls transition acts as the 
classical dynamic random field. The same type of behavior is typical for any
kind of soft -- mode transition. Damping effects naturally lead to finite 
correlation time of these fluctuations. 

Near a soft-mode (Peierls/CDW) transition this mode is usually overdamped as
$\gamma\gtrsim\omega_Q$. Resolved dynamic sidebands discussed in the main part of 
the text require $\omega_0\gg\Gamma_{\rm eff}$ i.e. 
$\omega_0^3\gg\gamma\Delta^2$, that is an underdamped mode whose frequency 
exceeds $(\gamma\Delta^2)^{1/3}$. Whether these criteria are satisfied in
real materials depends on the values of these parameters in concrete systems.


\end{document}